\documentclass[prx,a4paper,twocolumn,showpacs,floatfix,superscriptaddress,amsmath,amsfonts,amssymb,preprintnumbers,citeautoscript,aps,longbibliography]{revtex4-1}
\pdfoutput=1
\usepackage[T1]{fontenc}\usepackage[latin1]{inputenc}
\usepackage{dcolumn,graphicx,color,booktabs,microtype,afterpage} \graphicspath{{Figures/}}
\usepackage[charter,greekuppercase=italicized]{mathdesign}
\usepackage{sidecap}
\renewcommand{\figurename}{Fig.}
\renewcommand{\tablename}{Table}
\makeatletter\renewcommand{\fnum@figure}[1]{\figurename~\thefigure.}\makeatother
\makeatletter\renewcommand{\fnum@table}[1]{\tablename~\thetable.}\makeatother
\newcount\hh \newcount\mm
\hh=\time \divide\hh by 60
\mm=\hh \multiply\mm by 60 \mm=-\mm
\advance\mm by \time
\def\now{\number\hh:\ifnum\mm<10{}0\fi\number\mm}

\usepackage[colorlinks,plainpages=false,linkcolor=blue,urlcolor=blue,citecolor=blue,pdfpagemode=UseNone,pdfstartview=FitBH]{hyperref}

\newcommand{\McSymbol}{\includegraphics[height=6.8pt]{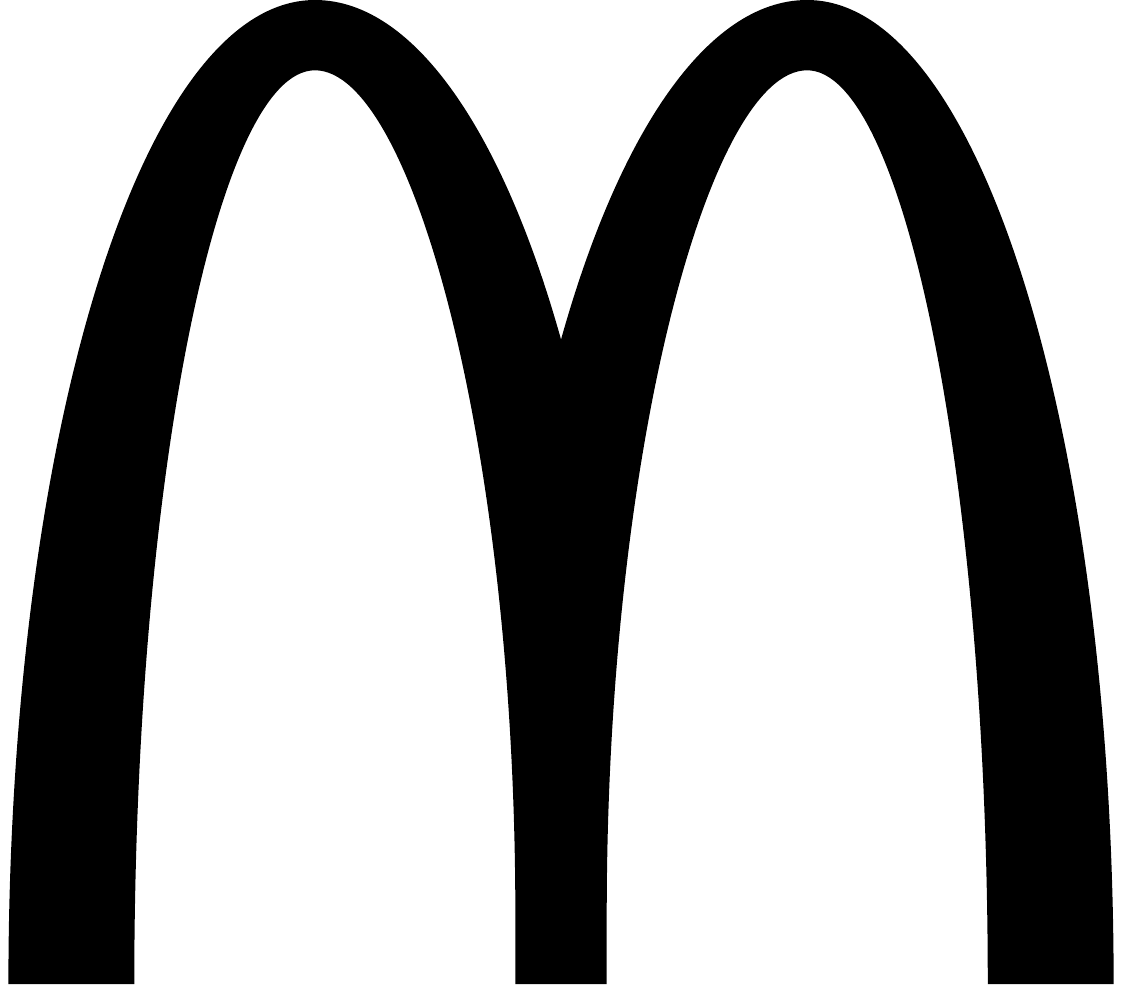}\kern0.5pt}

\begin{document}

\makeatletter\renewcommand{\ps@plain}{%
\def\@evenhead{\hfill\itshape\rightmark}%
\def\@oddhead{\itshape\leftmark\hfill}%
\renewcommand{\@evenfoot}{\hfill\small{--~\thepage~--}\hfill}%
\renewcommand{\@oddfoot}{\hfill\small{--~\thepage~--}\hfill}%
}\makeatother\pagestyle{plain}

%\preprint{\textit{Preprint: \today, \now. For internal use only, do not distribute.}}%\linenumbers

\title{Pseudo-Goldstone magnons in the frustrated \textit{S}\,=\,3/2 Heisenberg\\ helimagnet ZnCr$_\text{2}$Se$_\text{4}$ with a pyrochlore magnetic sublattice}

\author{Y.~V.~Tymoshenko}
\affiliation{Institut f\"ur Festk\"orper- und Materialphysik, Technische Universit\"at Dresden, D-01069 Dresden, Germany}

\author{Y.~A.~Onykiienko}
\affiliation{Institut f\"ur Festk\"orper- und Materialphysik, Technische Universit\"at Dresden, D-01069 Dresden, Germany}

\author{T.~M\"uller}
\affiliation{Institut f\"ur Theoretische Physik, Universit\"at W\"urzburg, 97074 W\"urzburg, Germany}

\author{R.~Thomale}
\affiliation{Institut f\"ur Theoretische Physik, Universit\"at W\"urzburg, 97074 W\"urzburg, Germany}

\author{S.~Rachel}
\affiliation{Institut f\"ur Theoretische Physik, Technische Universit\"at Dresden, D-01062 Dresden, Germany}
\affiliation{Department of Physics, Princeton University, Princeton, New Jersey 08544, USA}
\affiliation{School of Physics, University of Melbourne, Parkville, VIC 3010, Australia}

\author{A.~S.~Cameron}
\affiliation{Institut f\"ur Festk\"orper- und Materialphysik, Technische Universit\"at Dresden, D-01069 Dresden, Germany}

\author{P.~Y.~Portnichenko}
\affiliation{Institut f\"ur Festk\"orper- und Materialphysik, Technische Universit\"at Dresden, D-01069 Dresden, Germany}

\author{D.\,V.~Efremov}
\affiliation{Institute for Theoretical Solid State Physics, IFW Dresden, Helmholtzstra{\ss}e~20, D-01069 Dresden, Germany}

\author{V.~Tsurkan}
\affiliation{Experimental Physics V, Center for Electronic Correlations and Magnetism, Institute of Physics, University of Augsburg, D-86135 Augsburg, Germany}
\affiliation{Institute of Applied Physics, Academy of Sciences of Moldova, Chisinau MD-2028, Republic of Moldova}

\author{D.~L.~Abernathy}
\affiliation{Quantum Condensed Matter Division, Oak Ridge National Laboratory, Oak Ridge, Tennessee 37831, USA}

\author{J.~Ollivier}
\affiliation{Institut Laue-Langevin, 71 avenue des Martyrs, CS 20156, F-38042 Grenoble cedex 9, France}

\author{A.~Schneidewind}
\affiliation{J\"ulich Center for Neutron Science (JCNS), Forschungszentrum J\"ulich GmbH, Outstation at Heinz Maier\,--\,Leibnitz Zentrum~(MLZ), Lichtenberga{\ss}e~1, D-85747 Garching, Germany}

\author{A.~Piovano}
\affiliation{Institut Laue-Langevin, 71 avenue des Martyrs, CS 20156, F-38042 Grenoble cedex 9, France}

\author{V.~Felea}
\affiliation{Institute of Applied Physics, Academy of Sciences of Moldova, Chisinau MD-2028, Republic of Moldova}

\author{A.~Loidl}
\affiliation{Experimental Physics V, Center for Electronic Correlations and Magnetism, Institute of Physics, University of Augsburg, D-86135 Augsburg, Germany}

\author{D.~S.~Inosov}\email[Corresponding author: \vspace{-3pt}]{dmytro.inosov@tu-dresden.de}
\affiliation{Institut f\"ur Festk\"orper- und Materialphysik, Technische Universit\"at Dresden, D-01069 Dresden, Germany}

\begin{abstract}\parfillskip=0pt\relax%\linenumbers
\noindent Low-energy spin excitations in any long-range ordered magnetic system in the absence of magnetocrystalline anisotropy are gapless Goldstone modes emanating from the ordering wave vectors. In helimagnets, these modes hybridize into the so-called helimagnon excitations. Here we employ neutron spectroscopy supported by theoretical calculations to investigate the magnetic excitation spectrum of the isotropic Heisenberg helimagnet ZnCr$_2$Se$_4$ with a cubic spinel structure, in which spin-3/2 magnetic Cr$^\text{3+}$ ions are arranged in a geometrically frustrated pyrochlore sublattice. Apart from the conventional Goldstone mode emanating from the $(0\,0\,q_\text{h})$ ordering vector, low-energy magnetic excitations in the single-domain proper-screw spiral phase show soft helimagnon modes with a small energy gap of $\sim$\,0.17~meV, emerging from two orthogonal wave vectors $(q_\text{h}0\,0)$ and $(0\,q_\text{h}0)$ where no magnetic Bragg peaks are present. We term them pseudo-Goldstone magnons, as they appear gapless within linear spin-wave theory and only acquire a finite gap due to higher-order quantum-fluctuation corrections. Our results are likely universal for a broad class of symmetric helimagnets, opening up a new way of studying weak magnon-magnon interactions with accessible spectroscopic methods.
\end{abstract}


\maketitle

\section{\hspace{-1ex}Introduction}\label{Sec:Introduction}

\subsection{\hspace{-1ex}Classical Heisenberg model on the pyrochlore lattice}

The Heisenberg model on the pyrochlore lattice attracts a lot of interest as various spin models on this lattice give rise to the simplest three-dimensional frustrated spin systems. Even for classical spins, this model hosts a wide range of different ground states. Considering only the antiferromagnetic nearest-neighbor (NN) interactions results in a classical spin liquid \cite{MoessnerChalker98}, exhibiting no long-range magnetic order down to zero temperature. This is explained by strong geometric frustration that leads to a highly degenerate classical ground state. However, inclusion of further-neighbor interactions relieves this frustration and stabilizes various ordered ground states, among them ferromagnetism, single- or multi-$\mathbf{q}$ spin spirals, nematic order, and other exotic phases \cite{ReimersBerlinsky91, LapaHenley12, TakataMomoi15, OkuboNguyen11, ConlonChalker10}.

\begin{figure}[b]
\includegraphics[width=\linewidth]{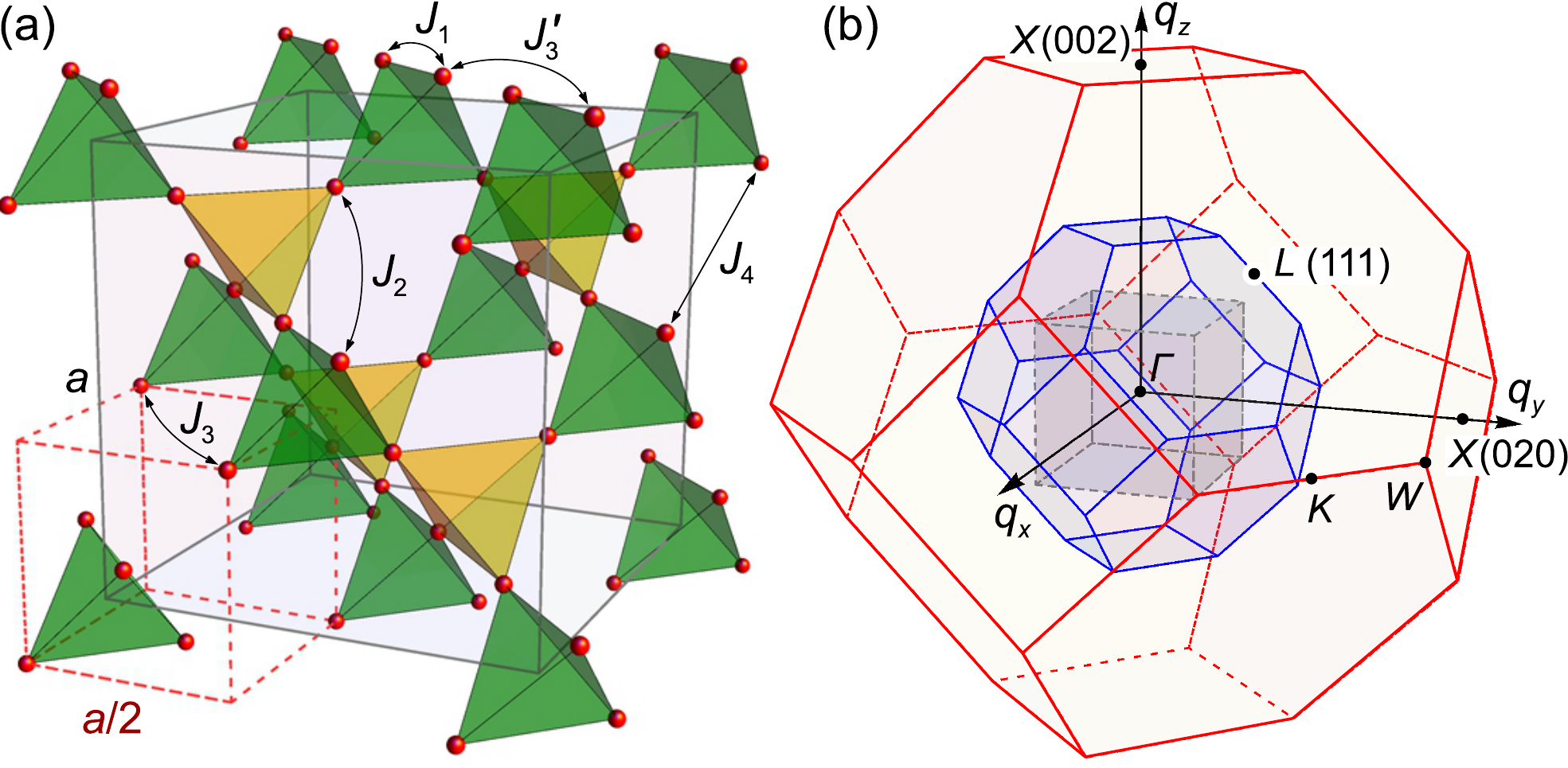}
\caption{~(a)~Pyrochlore-lattice structure, representative of the magnetic Cr sublattice in $A$Cr$_\text{2}X_\text{4}$ spinels. Small spheres represent Cr$^\text{3+}$ ions. The pyrochlore lattice can be described either as an fcc arrangement of separated Cr$_4$ tetrahedra formed by NN bonds (large cubic unit cell, solid lines) or as a half-filled fcc lattice of Cr$^\text{3+}$ ions with a twice smaller lattice parameter (small cubic unit cell, dashed lines). We also show different exchange paths corresponding to $J_1$\,...\,$J_4$ interactions that are discussed in the text. (b)~The simple-cubic Brillouin zone with dimensions $2\piup/a$ (central cube), and the two unfolded Brillouin zones corresponding to fcc lattices with parameters $a$ and $a/2$ (truncated octahedra). The high-symmetry points are marked according to the large unfolded zone.}
\label{Fig:PyrochloreStructure}
\end{figure}

Chromium spinels provide great opportunities to investigate magnetic interactions between classical spins on the structurally ideal pyrochlore lattice. These compounds have the general formula $A$Cr$_\text{2}X_\text{4}$, where $A$ and $X$ are non-magnetic ions and Cr$^\text{3+}$ is the magnetic cation in the $3d^3$ configuration \cite{Yaresko08}. Its magnetic sublattice has the pyrochlore structure with spins $S=3/2$ at the Cr sites. Using the classical Heisenberg model,
\begin{equation}
H=\sum_{ij} J_{ij} \mathbf{S}_i\cdot\mathbf{S}_j,\vspace{-4pt}
\end{equation}
is justified by the negligibly small magneto-crystalline anisotropy \cite{Siratori71, ZhangSu16, AoyamaKawamura16}. Thus we consider throughout the paper $J_{ij} \equiv J_{n}$ if sites $i$ and $j$ are $n^{\rm th}$ neighbors [see Fig.~\ref{Fig:PyrochloreStructure}\,(a)]. Depending on the chemical composition, chromium spinels exhibit different mechanisms of frustration, such as geometric frustration that occurs if dominant NN interactions are antiferromagnetic, or bond frustration which originates from competition between ferro\-magnetic NN and antiferromagnetic further-neighbor exchange.

To estimate the range and relative strengths of coupling constants $J_n$ in chromium spinels, Yaresko \cite{Yaresko08}  performed \textit{ab initio} calculations to extract exchange parameters up to the fourth nearest neighbor for various compounds of this family. Calculations showed that the NN interaction $J_1$ changes gradually from antiferromagnetic in some oxides to ferromagnetic in sulfides and selenides, while the next-nearest-neighbor (NNN) interaction $J_2$ is noticeably weaker than the antiferromagnetic $J_3$ exchange parameter (see Table~\ref{Table:ExchangeRatios}). For the HgCr$_2$O$_4$ system, $J_1$ can be even weaker than $J_2$ (or comparable, depending on the effective Coulomb repulsion $U$), so that the third-nearest-neighbor interaction $J_3$ may become dominant. Therefore, the existing theoretical phase diagram restricted to only NN and NNN interactions~\cite{LapaHenley12} appears insufficient for a realistic description of these materials. The importance of the two $3^\text{rd}$-nearest-neighbor exchange paths on the pyrochlore lattice has been also emphasized for the spin-$\frac{1}{2}$ molybdate Heisenberg antiferromagnet Lu$_2$Mo$_2$O$_5$N$_2$ \cite{ClarkNilsen14}, where $J_3'$ and $J_3''$ have opposite signs and dominate over $J_2$. It was recently conjectured that this may lead to an unusual ``gearwheel'' type of a quantum spin liquid \cite{IqbalMueller17}.

\begin{figure}[t]
\includegraphics[width=\linewidth]{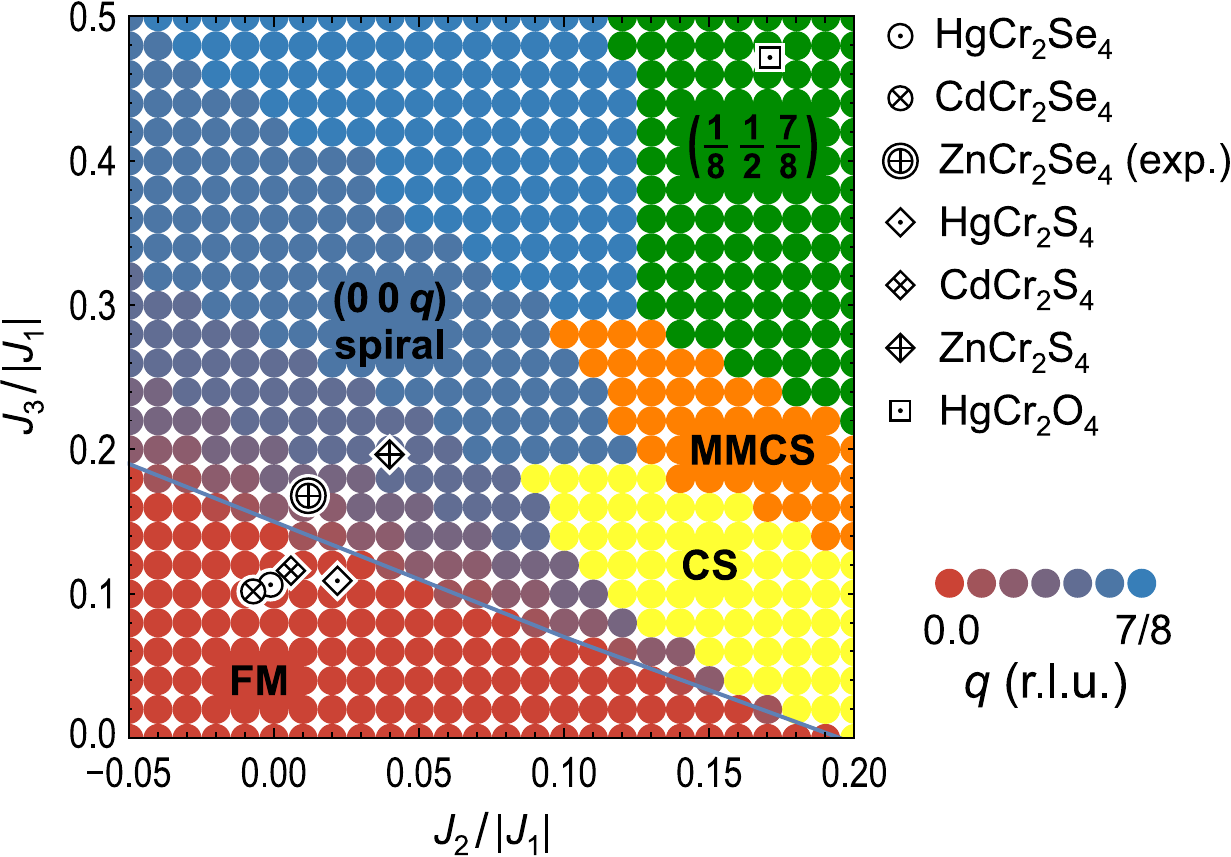}
\caption{~Classical $J_2$--$J_3$ phase diagram for $J_1=-1$ and $J_4/|J_1|=-0.014$ containing five phases: ferromagnetic (FM), $(00q)$ proper-screw spiral with the corresonding $q$-value shown in color code, the cuboctahedral stack (CS), the multiply modulated commensurate spiral (MMCS), and the $\left(\frac{1}{8}\frac{1}{2}\frac{7}{8}\right)$ phase. In addidtion, the location of the seven spinel compounds with ferromagnetic $J_1$ is indicated by the symbols. For details see the text and Table~\ref{Table:ExchangeRatios}.}
\label{Fig:PhaseDiagJ2J3}
\end{figure}

\vspace{-2pt}\subsection{\hspace{-1ex}Classical phase diagram relevant for chromium spinels}\vspace{-2pt}

An inspection of the theoretically predicted exchange parameters for various $A$Cr$_\text{2}X_\text{4}$ spinels presented in Ref.~\citenum{Yaresko08} reveals non-negligible exchange integrals up to the fourth-nearest neighbor, $J_4$, which is ferromagnetic in all the studied compounds. The third-nearest-neighbor exchange constants are, on the other hand, antiferromagnetic. Note that the pyrochlore lattice exhibits two inequivalent sets of third-nearest neighbors. Following Ref.~\citenum{Yaresko08}, we treat their corresponding exchange couplings as equal, $J_3'=J_3'' \equiv J_3$, since deviations are considered to be small. The classical ground state depends only on the ratios between the exchange constants, which can be much more accurately predicted in first-principles calculations than their absolute values. Here we restrict our attention to the sulfides, selenides, and HgCr$_2$O$_4$ with ferromagnetic NN interactions, $J_1<0$. In Table~\ref{Table:ExchangeRatios}, we summarize the calculated ratios $J_2/|\,J_1|$, $J_3/|\,J_1|$, and $J_4/|\,J_1|$ for all seven compounds considered in Ref.~\citenum{Yaresko08} (to be specific, we list the ratios for LSDA\,+\,$U$ with $U=2$~eV), along with their experimentally determined ground states.

We determine the classical phase diagram by means of a direct energy minimization scheme~\cite{LapaHenley12}. We consider a pyrochlore lattice (periodic boundary conditions imposed) with typically $L=8$ unit cells (containing 16 spins) in each spatial direction, resulting in $N=16 \cdot L^3=2^{13}$ spins in total. Starting from a random initial spin configuration, one randomly picks a lattice site $i$ and rotates its spin $\mathbf{S}_i$ antiparallel to its local field defined as
$
\mathbf{h}_i = \partial\!H/\partial\!\mathbf{S}_i = \sum_{j\not= i} J_{ij}\mathbf{S}_j,
$
thereby minimizing the energy and simultaneously keeping the spins normalized. Once an energetically minimal spin configuration has been found, the object of interest is the spin structure factor $\mathcal{S}(\mathbf{q}) = \frac{1}{N}\bigl|\sum_j \mathbf{S}_{\!j} \exp{({\rm i}{\kern.3pt}\mathbf{q}\!\cdot\!\mathbf{r}_{j})}\bigr|^2$ with its magnetic Bragg peaks.

\begin{table}[b!]
\begin{center}
\begin{tabular}{l@{~~}l@{~~}l@{~~}l@{~~}c@{~~}r}
    \toprule
        Compound & $J_2/|\,J_1|$ & $J_3/|\,J_1|$ & $J_4/|\,J_1|$ & ground state & Refs.\\
	\midrule
    	HgCr$_2$O$_4$ & 0.1714 & ~0.471 & ~\,0      & AFM & \cite{MatsudaUeda07}\\ \addlinespace[3pt]
        ZnCr$_2$S$_4$ & 0.0395 & ~0.198 & --0.014 & $(00q)$ spiral & \cite{YokaichiyaKrimmel09} \\
        CdCr$_2$S$_4$ & 0.0065 & ~0.116 & --0.015 & FM & \cite{MenyukDwight66} \\
        HgCr$_2$S$_4$ & 0.0222 & ~0.111 & --0.013 & $(00q)$ spiral & \cite{HastingsCorliss68, TsurkanHemberger06} \\ \addlinespace[3pt]
        ZnCr$_2$Se$_4$ & 0.0102 & ~0.169 & --0.018 & $(00q)$ spiral & \cite{Plumier66, CameronTymoshenko16}\\
        CdCr$_2$Se$_4$ & \!\!\!--0.0071 & ~0.101 & --0.013 & FM & \cite{MenyukDwight66}\\
        HgCr$_2$Se$_4$ & \!\!\!--0.0014 & ~0.109 & --0.013 & FM & \cite{Wojtowicz69}\\
    \midrule
        \multicolumn{6}{l}{Experimental values:}\\ \addlinespace[3pt]
        ZnCr$_2$Se$_4$ & 0.0118 & ~0.170 & --0.014 & \multicolumn{2}{r}{[this work]} \\
    \bottomrule
\end{tabular}
\caption{~Ratios of the exchange constants in different chromium spinels after Ref.\,\citenum{Yaresko08}, taken for a moderate effective Coulomb repulsion $U=2$\,eV. For each compound, its experimentally measured ground state is also given: antiferromagnetic (AFM), ferromagnetic (FM), or proper-screw spin spiral. The bottom row shows experimental values for ZnCr$_2$Se$_4$ estimated from our INS data in Section~II.B \mbox{of this work}.}\label{Table:ExchangeRatios}
\end{center}\vspace{-8pt}
\end{table}

The pure $J_1$--$J_2$ phase diagram already has been computed~\cite{LapaHenley12}. As mentioned before, we consider only ferromagnetic $J_1$. Table~\ref{Table:ExchangeRatios} reveals that the couplings $J_4$ of all considered spinels are ferromagnetic and roughly equal, $J_4/|J_1|\approx-0.014$ (which also corresponds to the value determined by fitting the experimental data to spin wave excitation spectra, see Sec.\,II.B). Thus we restrict our investigation of the phase diagram to $J_4/|J_1|=-0.014$, but we checked for all spinels that the ground state remains unchanged when taking into account the couplings listed in Table~\ref{Table:ExchangeRatios}. Therefore we map out the classical phase diagram for  $-0.05 \leq J_2/|J_1| \leq 0.2$ and $0 \leq J_3/|J_1| \leq 0.5$ as presented in Fig.\,\ref{Fig:PhaseDiagJ2J3}. We find five different phases: ferromagnetic (FM), $(00q)$ proper-screw spiral, cuboctahedral stack (CS), multiply modulated commensurate spiral (MMCS), and the $\bigl(\frac{1}{8} \frac{1}{2} \frac{7}{8}\bigr)$ phase. Quite generally,  we have observed for $J_2/|J_1| > 0.1$ and $J_3/|J_1| > 0.1$ that the energy landscape becomes extremely flat. In order to yield reliable results we had to increase system sizes drastically, for some parameter settings we even used $L=32$ corresponding to more than half a million spins. For dominant $J_1<0$ we find FM order, but both antiferromagnetic $J_2$ and $J_3$ destabilize the phase. For sufficiently strong $J_3>0$ (and not too large $J_2$) there is a phase transition from the FM to the $(00q)$ helical phase. The wave vector $q$ changes continuously from the phase transition line ($q=0$) up to a maximal value of $q=7/8$ for large $J_3$ (color-coded in increments of $q=1/8$ in Fig.\,\ref{Fig:PhaseDiagJ2J3}). For $J_3=0$ and $J_2/|J_1| \geq 0.2$ we find the CS phase of Ref.\,\cite{LapaHenley12} characterized by Bragg peaks at three of the four $\bigl\{\frac{1}{2} \frac{1}{2} \frac{1}{2}\bigr\}$-type $q$ vectors. Finite $J_3$ stabilizes the phase down to smaller values of $J_2$. Further increase of $J_3$ drives the system into the multi-$\mathbf{q}$ MMCS phase which is also present in a very small parameter range of the $J_1$--$J_2$ phase diagram\,\cite{LapaHenley12}.
This state is characterized by four main Bragg peaks at $\bigl(\frac{1}{4} \frac{3}{4} \frac{1}{2}\bigr)$, $\bigl(\frac{3}{4} \frac{1}{4} \frac{1}{2}\bigr)$, $\bigl(\frac{1}{4} \frac{3}{4} \frac{3}{2}\bigr)$ and $\bigl(\frac{3}{4} \frac{1}{4} \frac{3}{2}\bigr)$, and a subdominant peak at $\bigl(\frac{3}{4} \frac{3}{4}{\kern.3pt}0\bigr)$ (or symmetry related combinations of these vectors, respectively). Making $J_3$ even larger eventually turns the system's ground state into an exotic phase with four dominant Bragg peaks at $\mathbf{q}=\bigl(\pm\frac{1}{8}~\frac{1}{2}\,\pm\!\frac{7}{8}\bigr)$ and $\bigl(\pm\frac{7}{8}~\frac{1}{2}\,\pm\!\frac{1}{8}\bigr)$. To our best knowledge, this phase has not been mentioned before in the literature, and we simply call it the $\bigl(\frac{1}{8}\frac{1}{2}\frac{7}{8}\bigr)$ phase.

The phase diagram in Fig.\,\ref{Fig:PhaseDiagJ2J3} contains all spinels listed in Table~\ref{Table:ExchangeRatios}, and it is interesting to compare the results of our calculations with available experimental data. CdCr$_2$S$_4$, HgCr$_2$Se$_4$, and CdCr$_2$Se$_4$ have been reported to possess ferromagnetic ground states~\cite{MenyukDwight66, Wojtowicz69} in agreement with our simulations. HgCr$_2$S$_4$ develops a $(00q)$ spiral configuration with a temperature-dependent pitch \cite{HastingsCorliss68}, but the experimental results emphasize a strong tendency to ferromagnetic correlations~\cite{TsurkanHemberger06}; in our phase diagram it is located in the FM phase close to the spiral phase. The material of highest interest is, of course, ZnCr$_2$Se$_4$ where both experiment and simulation consistently find a $(00q)$ helix with $q=0.468$ (or 0.28\,\AA$^{-1}\!$). ZnCr$_2$S$_4$ has been reported to host a similar proper-screw spin structure in the temperature range $8~{\rm K} < T < 15$~K with $q=0.787$\,\cite{YokaichiyaKrimmel09}, which agrees fairly well with our simulated value $q=0.625$ based on the coupling constants from Table~\ref{Table:ExchangeRatios}. This spinel undergoes a structural phase transition at 8~K with a change of the magnetic order. The only oxide in our list is HgCr$_2$O$_4$, where Bragg peaks corresponding to $\mathbf{q}_1=(100)$ and $\mathbf{q}_2=(10\frac{1}{2})$ were measured\,\cite{MatsudaUeda07}. The compound undergoes, however, a structural phase transition into an orthorhombic phase\,\cite{UedaMitamura06} for which our pyrochlore description is of course inadequate. Otherwise, according to its exchange couplings, HgCr$_2$O$_4$ would be located in the $\bigl(\frac{1}{8} \frac{1}{2} \frac{7}{8}\bigr)$ phase.

We emphasize that our phase diagram, based on the {\it ab initio} parameters of Ref.\,\cite{Yaresko08}, agrees fairly well with the available experimental results. The major discrepancies are associated with structural transitions of the spinels ZnCr$_2$S$_4$ and HgCr$_2$O$_4$, leading to distortions or even different crystal structure. Since both effects cannot be captured by our simulations, the discrepancies between experiment and theoretical modeling in these particular cases are plausibly understood.

\vspace{-5pt}\subsection{\hspace{-1ex}Crystal structure and spin-spiral order in ZnCr$_2$Se$_4$}\vspace{-3pt}

Among the materials listed in Table~\ref{Table:ExchangeRatios}, characterized by ferromagnetic NN interactions, only ZnCr$_2$Se$_4$ and HgCr$_2$S$_4$ host an incommensurate spin-spiral ground state resulting from the bond frustration imposed by competition with further-neighbor interactions. The proper-screw spin structure of ZnCr$_2$Se$_4$ has a short helical pitch, $\lambda_\text{h}$, of only 22.4\,\AA\ \cite{Plumier66, AkimitsuSiratori78}, as compared to that of 42\,\AA\ in HgCr$_2$S$_4$ (which in addition increases with temperature to $\sim$90\,\AA\ at $T=30$\,K)~\cite{HastingsCorliss68}. This results in magnetic Bragg peaks in ZnCr$_2$Se$_4$ that are sufficiently remote from the commensurate structural reflection, with a propagation vector $(0~0~q_\text{h})$, where $q_\text{h} \approx 0.28$~\AA$^{-1}$~\cite{Plumier66, AkimitsuSiratori78}. Therefore, low-energy magnetic excitations emerging from different Bragg peaks are easy to resolve, which makes ZnCr$_2$Se$_4$ a perfect model material for investigations of low-energy spin dynamics in symmetric (Yoshimori-type) \cite{Yoshimori59, TogawaKousaka16} helimagnets that are common among multiferroics \cite{TokuraSeki10, Kimura12}. Moreover, from the comparison of the N\'{e}el temperature, $T_{\rm N}=21$\,K, with the Curie-Weiss temperature $\Theta_{\rm CW}=90$\,K, the frustration ratio of $\Theta_{\rm CW}/T_{\rm N}=4.3$ can be estimated, suggesting a considerable degree of frustration in this system \cite{RudolfKant07}. The small single-ion anisotropy energy in ZnCr$_2$Se$_4$ is evidenced by the magnetic resonance data~\cite{ZhangSu16}, by the gapless high-resolution neutron powder spectra down to at least 0.05~meV in energy \cite{ZajdelLi17}, as well as by the absence of any anisotropy-induced deformation of the spin spiral. According to a recent study \cite{AbdulJabbar15}, such a deformation would lead to the appearance of odd-integer higher-order magnetic Bragg peaks in neutron diffraction that are absent in the neutron-diffraction patterns of ZnCr$_2$Se$_4$ \cite{Plumier66, AkimitsuSiratori78}.

ZnCr$_2$Se$_4$ crystallizes in the normal spinel $(Fd\bar{3}m)$ structure at room temperature having 8 formula units per simple-cubic unit cell with a lattice constant ${\it{a}} = 10.497$~\AA~\cite{Plumier66, HidakaTokiwa03}. At low temperatures, the lattice undergoes a tiny distortion to an orthorhombic structure with $c/a=0.9999$ that could not be resolved in neutron diffraction measurements \cite{YokaichiyaKrimmel09}, but was suggested from earlier x-ray data \cite{HidakaTokiwa03, KleinbergerKouchkovsky66, HembergerNidda07}. For the purposes of our study, we neglect this distortion and describe the lattice as cubic at all temperatures. Cr$^{3+}$ ions form a pyrochlore magnetic sublattice which consists of corner-sharing tetrahedra [Fig.~\ref{Fig:PyrochloreStructure}\,(a)]. The sublattice can be described as a face-centered-cubic (fcc) arrangement of equally oriented Cr$_4$ tetrahedra (shown in dark color) or, alternatively, as a half-filled fcc lattice of individual Cr$^{3+}$ ions with a twice smaller unit cell, as shown in Fig.~\ref{Fig:PyrochloreStructure} with dashed lines. This allows us to introduce a larger unfolded Brillouin zone in reciprocal space, as shown in Fig.~\ref{Fig:PyrochloreStructure}\,(b), by analogy with the procedure described for the structurally related compound Cu$_2$OSeO$_3$ by Portnichenko \textit{et~al.} \cite{PortnichenkoRomhanyi16}.

\begin{figure*}[t]
\begin{center}\vspace{-1pt}
\includegraphics[width=\linewidth]{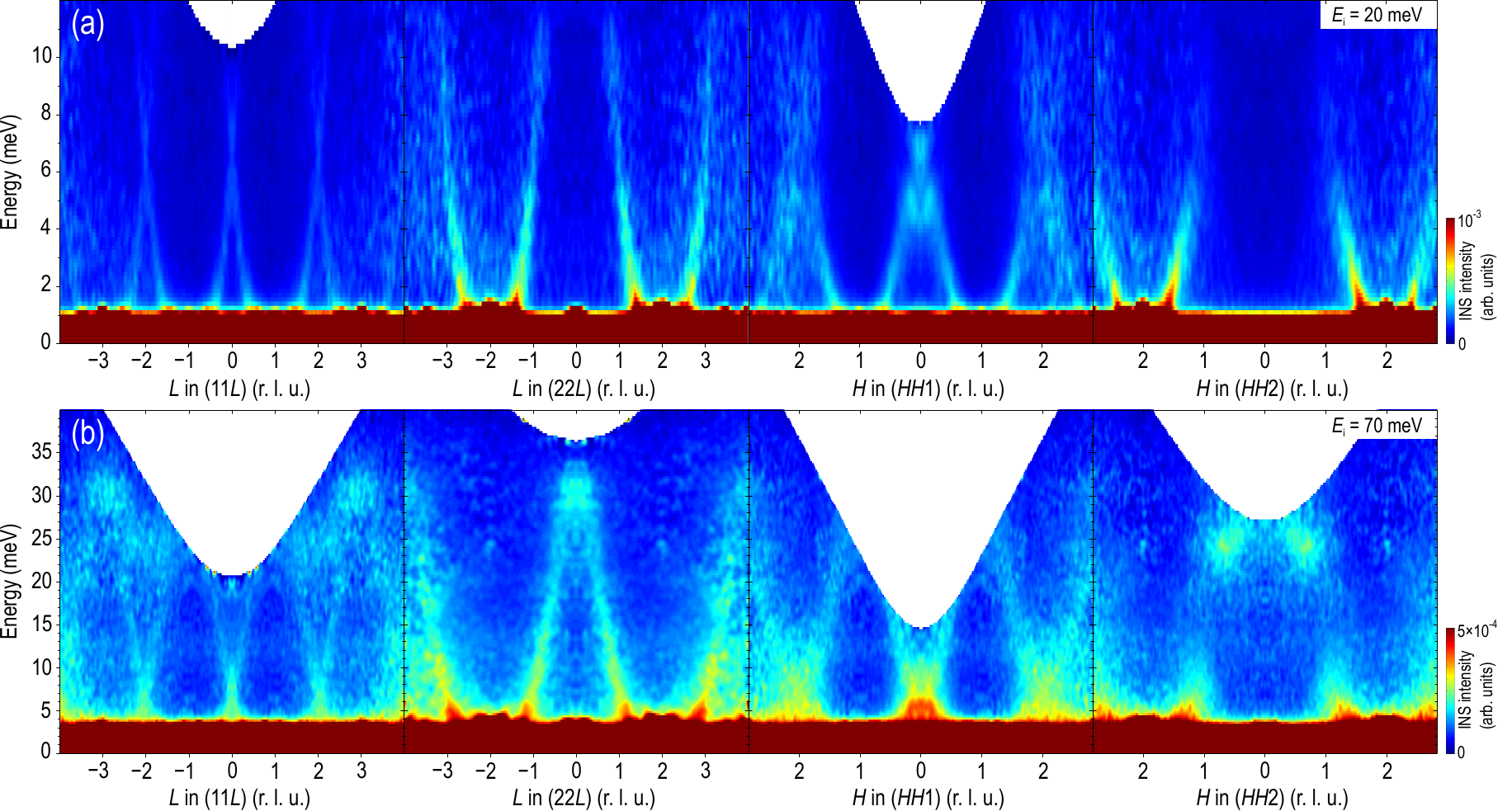}
\caption{~Energy-momentum cuts through the TOF data measured at the ARCS spectrometer (setup~1), plotted along high-symmetry directions after symmetrization respecting the cubic crystal symmetry. All the presented data were measured at the base temperature of 5\,K with incident neutron energies of 20~meV (top row) and 70~meV (bottom row). The momentum integration range in directions orthogonal to the image was set to $\pm$0.14~r.l.u. \vspace{-5pt}}\label{Fig:EnergyMomentumTOF}
\end{center}
\end{figure*}

Below the ordering temperature, Cr$^{3+}$ spins form an incommensurate helical structure with a propagation vector pointing along the [001] direction. In the absence of an external magnetic field, there are three possible domains with mutually perpendicular directions of the spin spirals, corresponding to the cubic crystal symmetry. Application of an external magnetic field leads to a domain selection transition, which stabilizes domains with the smallest angle between the propagation vector of the spiral and the magnetic field direction \cite{FeleaYasin12, CameronTymoshenko16}. This offers the possibility to prepare a single-domain state either by cooling the sample in magnetic field or by applying and removing the field at base temperature.

In this paper we present inelastic neutron scattering (INS) measurements of magnetic excitations in ZnCr$_2$Se$_4$ over a wide range of energies in the entire Brillouin zone. Comparing our data with spin-dynamical calculations, we have extracted exchange parameters up to the fourth nearest neighbor and found a good agreement with the isotropic Heisenberg model. Furthermore, measurements of low-energy helimagnon excitations performed in the single-domain spin-spiral state revealed two distinct modes: the Goldstone mode emanating from incommensurate magnetic Bragg peaks and a soft pseudo-Goldstone mode with a small spin gap of $\sim$0.17~meV that emerges from the structurally equivalent orthogonal wave vectors $(q_\text{h}\,0\,0)$ and $(0\,q_\text{h}\,0)$.

\vspace{-5pt}\section{\hspace{-1ex}Experimental results}\label{Sec:Exp_details}

\vspace{-3pt}\subsection{\hspace{-1ex}Instrumental conditions for the experiments}\vspace{-2pt}

We have used thermal- and cold-neutron time-of-flight (TOF) and triple-axis spectroscopy (TAS) techniques to map out dispersions of magnetic excitations in ZnCr$_2$Se$_4$ in a broad range of energies. The advantage of the TOF method is the possibility to obtain data covering the whole 4-dimensional (4D) energy-momentum space $(\mathbf{Q}, \hslash\omega)$ in a single measurement. Further, the combination of data collected with high-energy ($E_\text{i} = 20$ and 70 meV) neutrons at the ARCS spectrometer \cite{AbernathyStone12} at ORNL, Oak Ridge, and low-energy ($E_\text{i} = 3.27$~meV or $\lambda_\text{i}=5$\,\AA) neutrons at the IN5 spectrometer \cite{OllivierMutka11} at ILL, Grenoble, provides an overview of the whole magnon spectrum together with high-resolution imaging of low-energy excitations.
	
Experiments were performed on a mosaic of co-aligned single crystals of ZnCr$_2$Se$_4$ with a total mass $\sim$\,1\,g, prepared by chemical transport reactions and characterized as described elsewhere \cite{FeleaYasin12}. For the ARCS experiment (setup~1), the sample was mounted in such a way that the $(HHL)$ plane was horizontal. The experiment was performed without magnetic field in the multi-domain state at the base temperature of $T=5$\,K. We performed rocking scans for incoming neutron energies of 20 and 70~meV, which correspond to the energy resolution of 0.87 and 3.4~meV, respectively, defined as the full width at half maximum of the incoherent elastic line. We then processed the data using the \textsc{Horace} software package \cite{Horace, EwingsButs16}. The 4D datasets were symmetrized during data reduction by combining statistics from all symmetry-equivalent directions in the same Brillouin zone.

\begin{figure*}[t]
\begin{center}\vspace{-1pt}
\includegraphics[width=\linewidth]{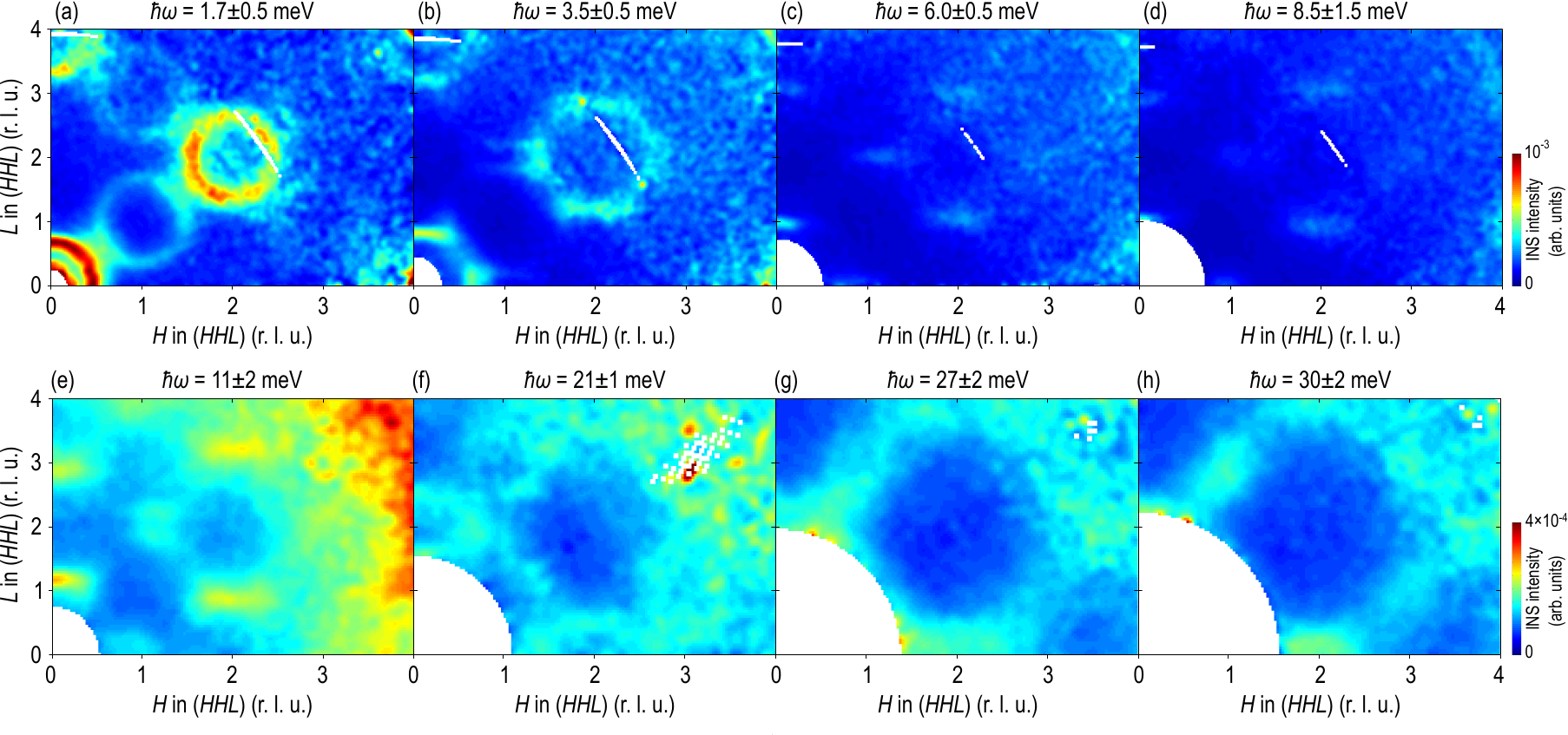}
\caption{~Constant-energy slices through the TOF data (setup~1) in the $(HHL)$ plane, integrated within different energy windows as indicated above each panel. The top and bottom rows of panels were measured at the base temperature of 5\,K with incident neutron energies of 20 and 70~meV. The out-of-plane momentum integration range was set to $\pm$0.1, $\pm$0.14, and $\pm$0.15~r.l.u. for panels (a), (b\,--\,d), and (e\,--\,h), respectively. The white stripes in some panels originate in regions with no data coverage.\vspace{-6pt}}\label{Fig:HHLmaps}
\end{center}
\end{figure*}

To confirm the positions of high-energy excitations at certain high-symmetry points of the Brillouin zone, we supplemented our TOF data with thermal-neutron TAS measurements carried out using the IN8 thermal-neutron spectrometer at ILL (setup~2). Here the sample was mounted inside an ``orange''-type cryostat in the $(HK0)$ scattering plane, so that the $W(320)$ point could be reached. The spectrometer was operated with the vertically focused pyrolytic graphite (PG) monochromator and analyzer, with the final neutron wave vector fixed to $k_{\rm f}=4.1$\,\AA$^{-1}$, and a PG filter installed between the sample and the analyzer. The collimation before and after the sample was set to 30$^\prime$ and $40^\prime$, respectively. The resulting energy resolution, calculated for this spectrometer configuration at 25~meV energy transfer, was about 4~meV.

The cold-neutron TOF experiment at IN5, ILL (setup~3), was carried out using the same sample placed in the ``orange''-type 2.5\,T cryomagnet. This time the sample was rotated about the $(111)$ axis so that its $(11\overline{2})$ direction was pointing vertically (parallel to the direction of the magnetic field). Correspondingly, the equatorial scattering plane was spanned by the mutually orthogonal $(1\overline{1}0)$ and $(111)$ vectors. In order to stabilize one helimagnetic domain, we cooled down the sample in a vertical magnetic field of 1.5~T, with $\mathbf{B}\parallel(11\overline{2})$. In this geometry, the $(0\,0\,q_\text{h})$ ordering vector forms an angle of 35.3$^\circ$ with the field direction, leading to a twice larger projection of the field on this axis as compared to the ordering vectors of the two other domains, $(0\,q_\text{h}0)$ and $(q_\text{h}0\,0)$, that are inclined at 65.9$^\circ$ with respect to the magnetic field. As a consequence, out of the three possible helimagnetic domains, only the one with the propagation vector along the $(001)$ direction is selected by the applied field. When the base temperature of 1.5\,K was reached, the magnetic field was switched off, and the measurement was performed in the single-domain state in zero field with the incident neutron wavelength of 5\,\AA\ and the elastic energy resolution of 0.084~meV.
	
In this scattering geometry, neither of the three propagation vectors lies within the horizontal scattering plane. However, they are sufficiently short to be covered by the out-of-plane acceptance range of the position-sensitive TOF detector bank despite the limitation from the vertical \mbox{opening} angle of the cryomagnet. Therefore, magnetic satellites of the $(000)$, $(111)$, and $(222)$ zone centers could be accessed in this configuration, which would be unfeasible on most instruments restricted to the horizontal scattering plane. During data reduction, the complete dataset was then converted into energy-momentum space using the conventional basis vectors of the simple-cubic crystallographic unit cell, so that the presentation of the data is not affected by the rotated sample mounting. We visualize the obtained results by presenting two-dimensional (2D) cuts through the 4D dataset that are taken along high-symmetry momentum directions. To designate points in $\mathbf{Q}$ space, we will use reciprocal lattice units corresponding to the simple-cubic unit cell ($1\,\text{r.l.u.} = 2\pi/a$), whereas high-symmetry points will be marked according to the unfolded Brillouin zone following the standard notation for an fcc lattice \cite{PortnichenkoRomhanyi16}.

Finally, we performed detailed spin-gap measurements at the cold-neutron TAS \textsc{Panda} \cite{SchneidewindCermak15} operated by JCNS at MLZ, Garching (setup~4). They were also carried out at zero field in the single-domain state, which was prepared by field-cooling the sample in a field of 2.5\,T using the 5\,T vertical-field cryomagnet. The same mosaic sample was remounted with its $[001]$ axis pointing vertically, along the magnetic field direction, which resulted in the selection of the domain with the $(0\,0\,q_\text{h})$ ordering vector, whereas the accessible range of $\mathbf{Q}$ space was restricted to the $(H\,K\,0)$ scattering plane spanned by the propagation vectors of the two suppressed domains, $(q_\text{h}0\,0)$ and $(0\,q_\text{h}0)$. Their Bragg intensities were reduced by a factor of $\sim$\,200 as a result of domain selection. To resolve these wave vectors at low scattering angles (in the vicinity of the direct beam), we operated the spectrometer with the horizontally flat but vertically focused monochromator and analyzer. The final neutron energy was fixed at $E_\text{f}=3.0$~meV ($k_\text{f}=1.2$\,\AA$^{-1}$), providing an energy resolution of about 0.1~meV. A cold beryllium filter was used to suppress higher-order contamination from the monochromator.
		
\vspace{-6pt}\subsection{\hspace{-1ex}Experimental determination of the exchange constants}\vspace{-3pt}

We will start the presentation of our results with the higher-energy data measured at the ARCS spectrometer in the multi-domain state. Figure~\ref{Fig:EnergyMomentumTOF} shows typical energy-momentum cuts along several high-symmetry lines in $\mathbf{Q}$ space that are parallel to the (001) and (110) directions. Panels (a) and (b) show low- and high-energy datasets ($E_\text{i} = 20$ and 70 meV), respectively. In the $(22L)$ and $(HH2)$ cuts, we clearly see intense magnon branches emanating from incommensurate magnetic satellites of the $(222)$ Bragg peak, while similar modes near the $(111)$ and $(113)$ points are noticeably weaker. This behavior of the dynamic structure factors can be explained using the approach of unfolded Brillouin zones, which was detailed in Ref.~\citenum{PortnichenkoRomhanyi16} for another compound with a structurally similar magnetic sublattice. One characteristic energy scale revealed in Fig.~\ref{Fig:EnergyMomentumTOF}\,(a) is given by the crossing point between equivalent magnon branches in the $(11L)$ direction, located at $\sim$\,6.8~meV.

\begin{figure}[t!]
\includegraphics[width=\linewidth]{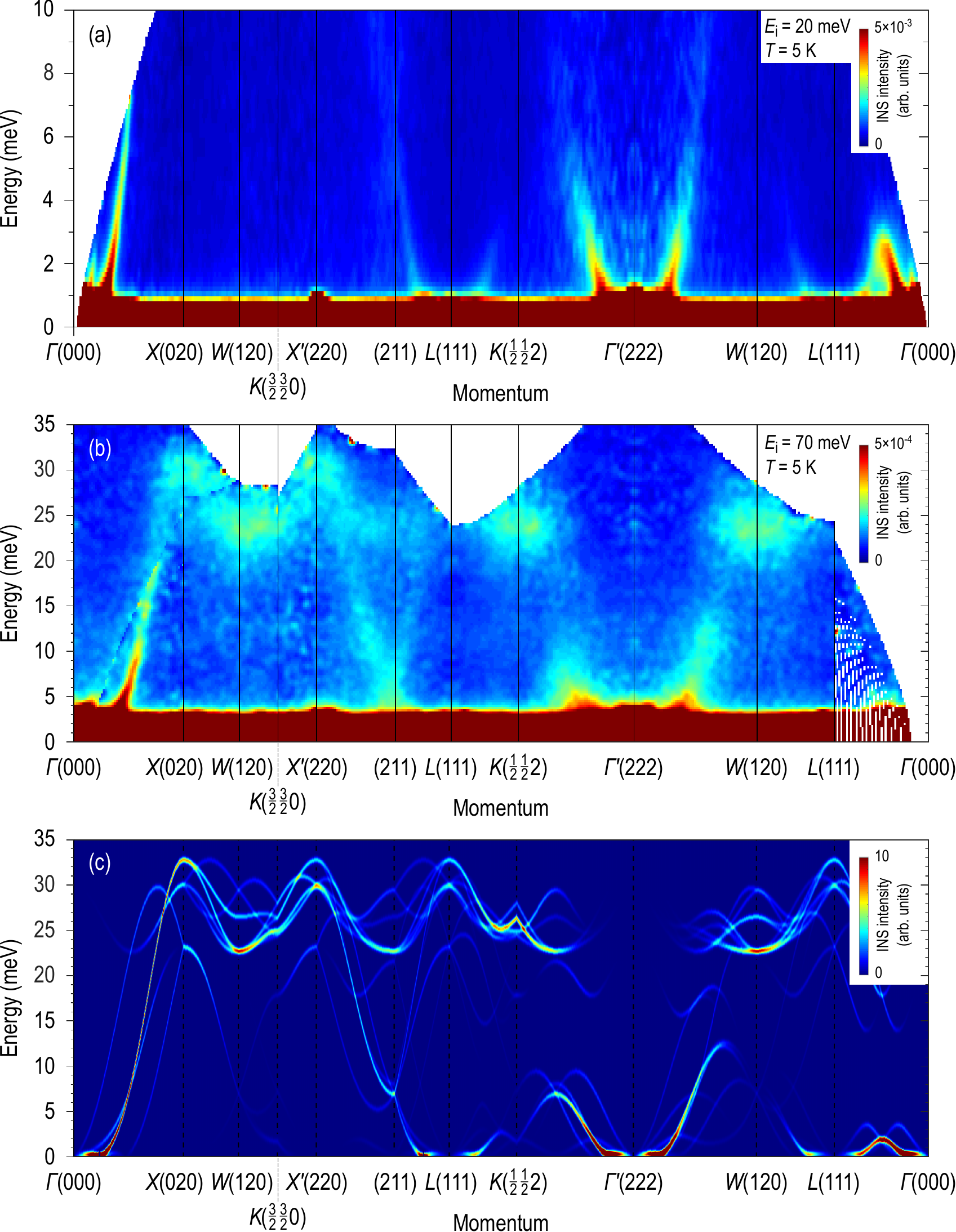}
\caption{~(a,b) Energy-momentum cuts thought the TOF data, taken with $E_\text{i} = 20$ and 70~meV at the ARCS spectrometer (setup~1), combined into a ``spaghetti''-type plot that covers all high-symmetry directions of the unfolded Brillouin zone. (c)~Results of our spin-dynamical calculations performed using the \textsc{SpinW} software for the exchange constants given in Eq.~\ref{Eq:ExchangeConstants}, plotted along the same polygonal path in momentum space as the data in panels (a) and (b). The calculations were averaged over three possible domain orientations to mimic the multi-domain spin-spiral state realized in a zero-field experiment.\vspace{-3pt}}\label{Fig:Spaghetti}
\end{figure}

In Fig.~\ref{Fig:EnergyMomentumTOF}\,(b), we additionally observe high-energy modes near the top of the magnon band, between 25 and 35~meV. While there is no energy gap separating the high- and low-energy modes, unlike in Cu$_\text{2}$OSeO$_\text{3}$ \cite{PortnichenkoRomhanyi16}, the intensity is certainly enhanced near the top of the magnon band, as best seen along the $(HH2)$ direction (rightmost panel). This is reminiscent of the situation in another recently studied helimagnet with a similar pitch of the spin spiral, Sr$_3$Fe$_2$O$_7$, where an intense and well separated high-energy magnon branch was observed at a similar energy of 25~meV \cite{KimJain14}. However, the low-energy behavior of the spin waves in Sr$_3$Fe$_2$O$_7$ is considerably different. There, apart from the steeply dispersing outward branches, we observed \McSymbol-shaped inner branches of helimagnetic spin waves connecting the incommensurate ordering vectors, which extended up to $\sim$\,4~meV in energy. This form of spin-wave dispersion, bridging the incommensurate magnetic satellites, is a common feature of helimagnets with symmetric exchange interactions \cite{TogawaKousaka16}. In the case of ZnCr$_2$Se$_4$, however, this branch occurs at 10 times lower energies, so it cannot be seen in Fig.~\ref{Fig:EnergyMomentumTOF}\,(a) above the elastic line. We will return to the discussion of this low-energy magnon branch when presenting the cold-neutron data in Section~II.C.

To visualize the magnon spectrum in $\mathbf{Q}$ space, we are presenting constant-energy cuts in Fig.~\ref{Fig:HHLmaps}. All of them show intensity distributions within the $(HHL)$ plane at different energies, covering wave vectors up to $(444)$. Here one can observe the complex hierarchy of energy scales in the magnon dispersion. The low-energy cut at 1.7~meV in Fig.~\ref{Fig:HHLmaps}\,(a) displays clearly separated rings of scattering intensity around the zone centers. Stronger modes are found near $\Gamma$ points of the unfolded Brillouin zones, such as $\Gamma(000)$, $\Gamma(222)$, and $\Gamma(004)$, whereas weaker modes appear as replicas shifted by the $(111)$ wave vector to the centers of the folded fcc Brillouin zone \cite{PortnichenkoRomhanyi16}. Around 2.5~meV, these two modes start to hybridize, leading to a saddle point in the spin-wave dispersion that can be most clearly seen along the $(HHH)$ direction at the right-hand side of Fig.~\ref{Fig:Spaghetti}\,(a). Above the saddle-point energy, the rings of scattering break into separated segments as shown in Figs.~\ref{Fig:HHLmaps}\,(b\,--\,e). Above 6.8~meV, which corresponds to the crossing of the bands in the $(11L)$ direction in Fig.~\ref{Fig:EnergyMomentumTOF}\,(a), the weaker rings of scattering surrounding the $(111)$ and $(113)$ points cross each other, resulting in a streak of intensity centered at $(112)$ in Figs.~\ref{Fig:HHLmaps}\,(c,d). Going even higher in energy, we observe an elliptical feature surrounding the $X(002)$ point in Fig.~\ref{Fig:HHLmaps}\,(f) that finally shrinks in Figs.~\ref{Fig:HHLmaps}\,(g,h) into a set of broad peaks corresponding to the top of the magnon band at the unfolded-zone boundary at an energy of $\sim$\,30~meV. What is not seen in this figure is the bottom of the high-energy magnon branch that results in an intense peak centered at 24~meV at the $W(120)$ point, which lies outside of the $(HHL)$ plane, but can be well seen in Fig.~\ref{Fig:Spaghetti}\,(b). A similar feature was also seen in the spectrum of Cu$_\text{2}$OSeO$_\text{3}$ at the same $\mathbf{Q}$ point and nearly the same energy \cite{PortnichenkoRomhanyi16}.

\begin{figure}[b!]
\includegraphics[width=\linewidth]{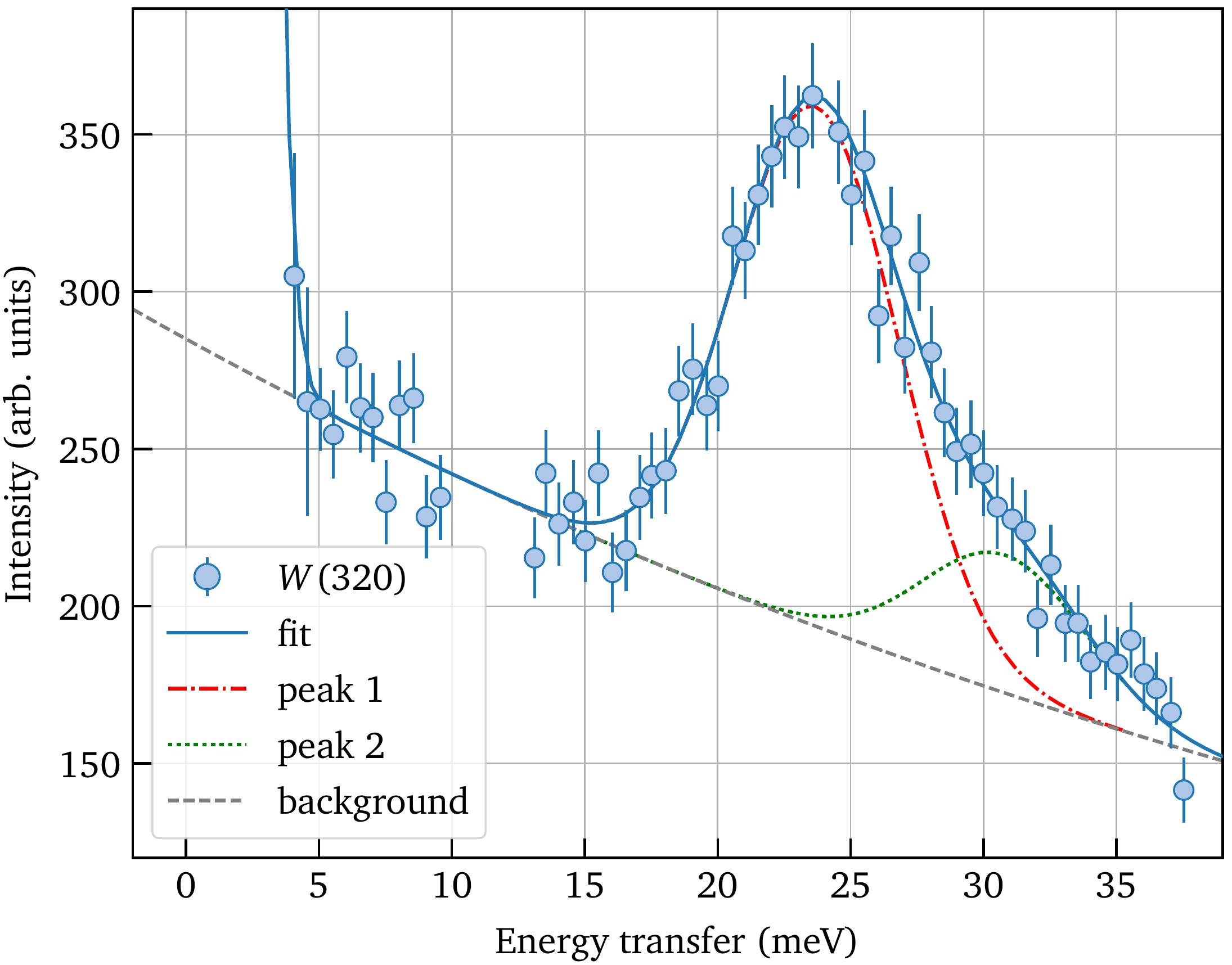}\vspace{-2pt}
\caption{~A typical energy scan measured at the $W(320)$ point with the IN8 spectrometer (setup~2) in the multi-domain state, revealing two marginally resolved peaks centered at $23.7$~meV (dash-dotted line) and $30.8$~meV (dotted line), in perfect agreement with our spin-dynamical calculations.\vspace{-5pt}}\label{Fig:IN8_W}
\end{figure}
	
\begin{figure*}[t!]
\includegraphics[width=\linewidth]{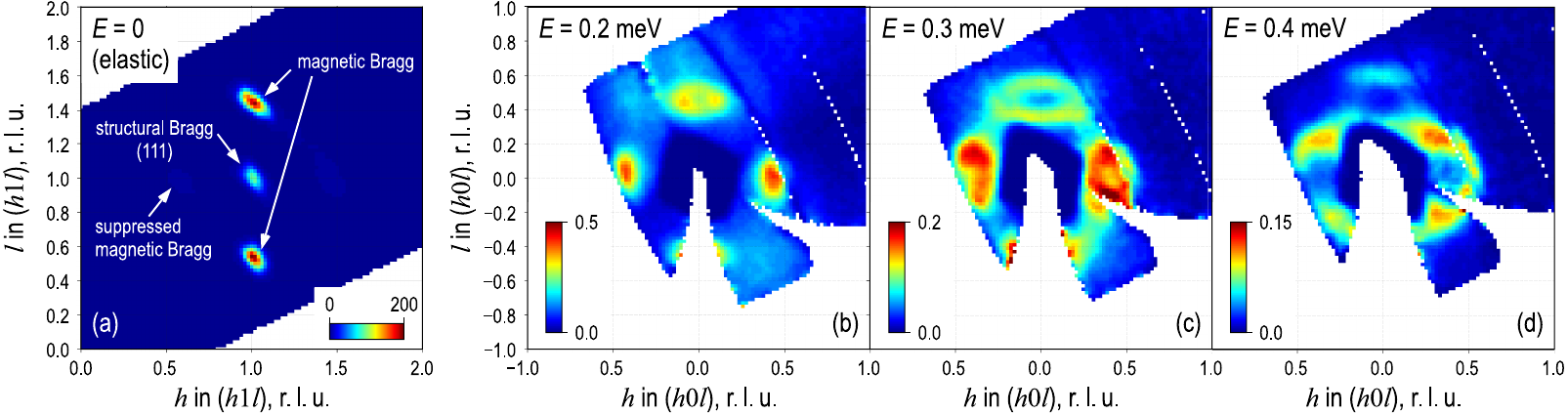}
\caption{~Constant-energy cuts through the low-energy TOF data measured at IN5 (setup~3) in the single-domain state. (a)~Elastic intensity map in the $(H1L)$ plane, demonstrating the selection of a single magnetic domain with the propagation vector parallel to the $\mathbf{c}$ axis. (b\,--\,d)~Constant-energy cuts at 0.2, 0.3, and 0.4~meV, which show magnetic Goldstone modes emanating from the $(0~0\,\pm\!q_\text{h})$ ordering vectors and pseudo-Goldstone modes at the orthogonal $(\pm q_\text{h}\,0\,0)$ vectors, which are seen to merge at 0.4~meV. Note different intensity scales in all images.}
\label{Fig:MapsIN5}
\end{figure*}
\begin{figure*}[t!]
\includegraphics[width=\linewidth]{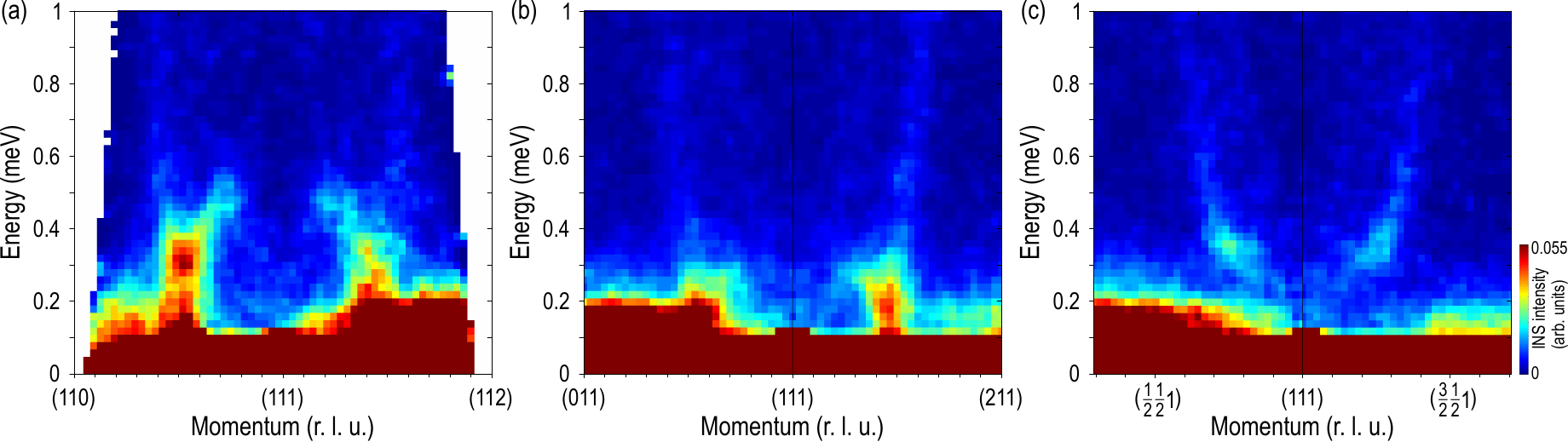}
\caption{~Energy-momentum cuts through the IN5 data (setup~3), centered at the $(111)$ wave vector and taken along different directions: (a)~along the $(11L)$ line passing through the $(1\,1\,1\!\pm{\kern-0.6pt}q_\text{h})$ ordering vectors of the field-selected domain; (b)~average of the equivalent $(1\!\pm{\kern-0.6pt}h\,1\,1)$ and $(1\,1\!\pm{\kern-0.6pt}h\,1)$ directions passing through the suppressed $(1\!\pm{\kern-0.6pt}q_\text{h}\,1\,1)$ and $(1\,1\!\pm{\kern-0.6pt}q_\text{h}\,1)$ ordering vectors; (c)~along two orthogonal diagonal directions: $(1\!-{\kern-0.6pt}h\,1\!-{\kern-0.6pt}h\,1)$ (left) and $(1\!\pm{\kern-0.6pt}h\,1\!\mp\!h\,1)$ (right), which pass through the 0.35~meV inflection point in the dispersion in the plane orthogonal to the ordering vector. In the right part of panel (c), datasets for positive and negative $h$ have been averaged.\vspace{-3pt}}
\label{Fig:EnergyMomentumIN5}
\end{figure*}

For a more comprehensive representation of magnetic excitations in ZnCr$_2$Se$_4$, we put together energy-momentum cuts along a polygonal path which contains all high symmetry directions in $\mathbf{Q}$ space, creating a ``spaghetti''-type plot shown in Figs.~\ref{Fig:Spaghetti}\,(a,b) for both incident energies. Kinematic constraints lead to the lack of data at high energy transfer in the first Brillouin zone, however these data have better signal-to-noise ratio at low energies. Therefore, in Fig.~\ref{Fig:Spaghetti}\,(b) we have overlayed data from the first Brillouin zone on top of those obtained from equivalent points at higher $|\mathbf{Q}|$ to complete the missing parts of the dataset. The stitching lines between high- and low-$|\mathbf{Q}|$ data can be recognized at the left-hand side of the figure.

To extract the experimental exchange constants from our data, we compared them with spin-dynamical calculations performed in the framework of linear spin-wave theory (LSWT) using the \textsc{SpinW} software package \cite{SpinW, TothLake15}. The magnon spectrum was calculated using the classical Heisenberg model with interactions up to the 4$^\text{th}$ shell of Cr neighbors \cite{Yaresko08}:\vspace{-5pt}
\begin{equation}\label{Eq:HeisenbergModel}
H = \frac{1}{4}\sum_{i=1}^4\sum_{n=1}^4\sum_{j=1}^{z_n} J_n\,\mathbf{S}_i \cdot \mathbf{S}_j,
\end{equation}
where $i$ numbers Cr sites in the unit cell, and $j$ runs over $z_n$ neighbors in the $n^\text{th}$ coordination shell around the site~$i$. In our notation, positive $J_n$ stands for AFM coupling between Cr $S=3/2$ spins. The model does not differentiate between two inequivalent exchange paths for third-nearest neighbors, $J_3^\prime$ and $J_3^{\prime\prime}$, hence these two exchange parameters were assumed equal. The pitch angle $\gamma$ between the neighboring magnetic moments of the spin helix can be found by solving the equation that minimizes the energy for classical spins:
\begin{equation}
J_1 + J_2+4J_4 + 4(J_2+2J_3)\cos(\gamma) + 6 J_4 \cos(2\gamma)=0.
\end{equation}

To enable direct comparison with our ARCS data, we modeled the multi-domain state by averaging the calculated spectra for three possible orientations of the magnetic domains. The calculations were first carried out with the theoretically predicted exchange parameters from Ref.~\citenum{Yaresko08}, which were then iteratively adjusted to achieve the best agreement with the measured spectrum. The resulting spectrum is shown in Fig.\,\ref{Fig:Spaghetti}\,(c) for the following optimized values of exchange constants:
\begin{equation}\label{Eq:ExchangeConstants}
\begin{aligned}
J_1&= -2.876\,\text{meV,} & J_2&=0.034\,\text{meV,}\\
J_3^\prime&=J_3^{\prime\prime}=0.490\,\text{meV,} & J_4^{\phantom{\prime}}&=-0.041\,\text{meV.}
\end{aligned}
\end{equation}
The absolute value of the experimental NN exchange constant $J_1$ is approximately 50\% lower than the theoretical prediction. However, the signs of all interactions are well captured by the calculation. Their ratios, listed in Table~\ref{Table:ExchangeRatios}, show fair agreement (within 20\%) with the results obtained in Ref.~\citenum{Yaresko08} for an effective Coulomb repulsion $U=2$\,eV.

As an additional verification of the exchange parameters and the overall correctness of our spin-wave model, in Fig.~\ref{Fig:IN8_W} we show a typical energy scan measured at the IN8 thermal-neutron spectrometer at the $W(320)$ wave vector. It reveals an intense peak corresponding to the bottom of the high-energy magnon branch (cf. Fig.~\ref{Fig:Spaghetti}). A closer analysis of the peak shape reveals that it consists of a stronger excitation centered at $(23.7\pm0.2)$~meV and a weaker one at $(30.8\pm1.2)$~meV. The strong lower-energy peak originates from the sum of two magnetic domains whose propagation vectors are directed along the $\mathbf{y}$ and $\mathbf{z}$ cubic axes, whereas the weaker upper peak presumably consists of two unresolved excitations originating from the $\mathbf{x}$-oriented domains. Both the positions and the relative intensities of the two peaks match fairly well with the spin-dynamical calculation results in Fig.~\ref{Fig:Spaghetti}\,(c).

\vspace{-5pt}\subsection{\hspace{-1ex}TOF measurements in the single-domain state}\vspace{-3pt}

We proceed with presenting low-energy data measured at the IN5 spectrometer in the single-domain state prepared by cooling the sample in magnetic field as described in Section II.A. To confirm that our domain-selection procedure was successful, in Fig.~\ref{Fig:MapsIN5}\,(a) we show the elastic-scattering intensity map in the $(h1l)$ plane in the vicinity of the $(111)$ structural Bragg reflection. This plane contains both the $(1~1~1\!\pm\!q_\text{h})$ ordering vectors, where we observe intense magnetic Bragg peaks, and the $(1\!\pm\!q_\text{h}~1~1)$ wave vectors, where the corresponding peaks are suppressed by two orders of magnitude in intensity. We therefore conclude that all the inelastic spectra obtained from the same dataset originate from a single magnetic domain oriented along the $\mathbf{c}$ axis.

\begin{figure}[t!]\vspace{3pt}
\includegraphics[width=\linewidth]{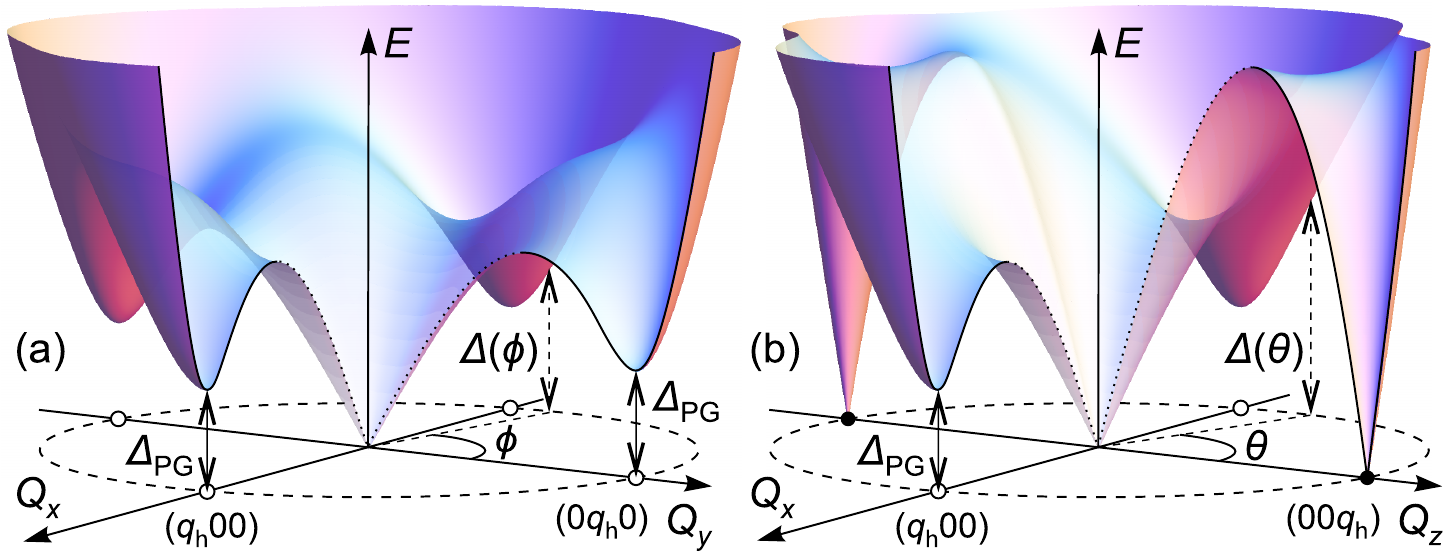}\vspace{-2pt}
\caption{~Schematic illustration of the low-energy magnon dispersion: (a) in the $(HK0)$ plane orthogonal to the ordering vector and (b)~in the $(H0L)$ plane passing through the ordering vector. Here $\Delta_{\rm PG}$ denotes the gap of the pseudo-Goldstone magnon branch, $\phi$ and $\theta$ are spherical angles. The filled and open dots represent Bragg peaks of the selected and suppressed magnetic domains, respectively.\vspace{-3pt}}\label{Fig:MexicanHat}
\end{figure}

The inelastic data shown in Fig.~\ref{Fig:MapsIN5}\,(b\,--\,d) represent equivalent cuts taken in the vicinity of $\mathbf{Q}=0$, where the signal-to-background ratio is maximized. The constant-energy maps exhibit two inequivalent types of low-energy spin-wave modes seen as elliptical features emanating from the selected and suppressed ordering vectors. From different sizes of the ellipses it is clear that the dispersion of the two modes is not identical. Moreover, the intensity of the second mode corresponding to the suppressed Bragg reflection appears to be even higher compared to the true Goldstone mode. The two modes merge at 0.4\,meV [Fig.~\ref{Fig:MapsIN5}\,(d)], resulting in an inflection point (flattening) of the dispersion seen as bright intensity spots along the \mbox{diagonal directions of the image}.

The dispersion of the same low-energy excitations can be seen in Fig.~\ref{Fig:EnergyMomentumIN5}, where we compare energy-momentum cuts passing through the $(1\,1\,1\!\pm{\kern-0.6pt}q_\text{h})$ ordering vectors with orthogonal cuts through the suppressed magnetic Bragg positions [panels (a) and (b), respectively]. In both directions, we observe Goldstone-like magnon modes that appear gapless within the experimental energy resolution. Their dispersion resembles the \McSymbol-shaped low-energy magnon branch in Sr$_3$Fe$_2$O$_7$ \cite{KimJain14}, with the top of the inner branch reaching to only 0.5~meV for the Goldstone modes in Fig.~\ref{Fig:EnergyMomentumIN5}\,(a) and 0.28~meV for the pseudo-Goldstone modes in Fig.~\ref{Fig:EnergyMomentumIN5}\,(b). A weak spin-wave band dispersing downward towards $(111)$ can be seen in both directions. The nearly twofold difference in the band width of the inner \McSymbol-shaped branch connecting the magnetic Bragg peak with the zone center confirms that the two modes seen in panels (a) and (b) are distinct. The behavior of the steeply dispersing outward branch is, on the other hand, much more isotropic. We also show in Fig.~\ref{Fig:EnergyMomentumIN5}\,(c) diagonal cuts that cross the inflection points of the spin-wave dispersion in the plane perpendicular to the ordering vector. As can be estimated from the image, these inflection points are located at 0.35~meV in energy.

\vspace{-3pt}\subsection{\hspace{-1ex}Spin gap of the pseudo-Goldstone magnons}\vspace{-3pt}

\begin{figure}[b!]
\includegraphics[width=\linewidth]{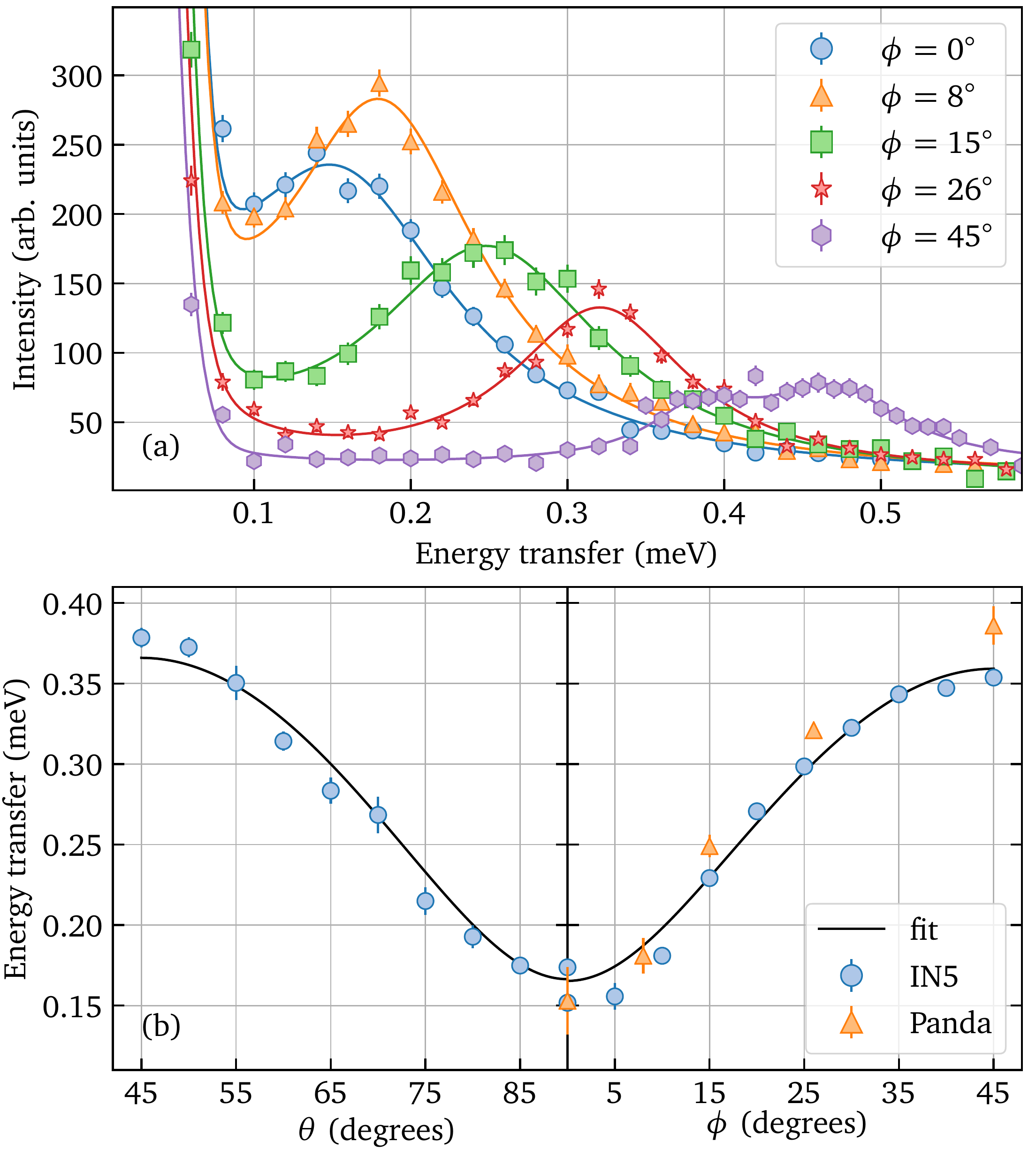}\vspace{-2pt}
\caption{~(a)~Several unprocessed energy scans from the triple-axis spectrometer \textsc{Panda} (setup~4), measured at the minimum of the spin-wave dispersion in the plane orthogonal to the ordering vector ($\theta=\pi/2$) along different directions with $\phi=0^\circ$, 8$^\circ$, 15$^\circ$, 26$^\circ$, and 45$^\circ$. (b)~Angular dependence of the minimum energy in the spin-wave dispersion $\Delta(\theta,0)$ and $\Delta(\pi/2,\phi)$ extracted from the IN5 and \textsc{Panda} measurements. The solid line is an empirical fit described in the text.\vspace{-3pt}}\label{Fig:ExperimentGap}
\end{figure}

The presented IN5 data indicate that a soft magnon mode persists practically in all momentum directions around the zone center at $|\mathbf{Q}|\approx q_\text{h}$, by analogy with the result obtained by Kataoka \cite{Kataoka87} for conical spin density waves with symmetric exchange interactions. This results in a corrugated Mexican-hat-like dispersion schematically shown in Fig.~\ref{Fig:MexicanHat}. The locus of dispersion minima along different directions forms an approximately spherical surface in $\mathbf{Q}$ space of radius $q_\text{h}$. On this surface, the spin gap $\Delta(\theta,\phi)$ has two distinct minima: at the ordering vector itself ($\theta=0$), forming the true (gapless) Goldstone mode emanating from the magnetic Bragg peaks (filled dots), and pseudo-Goldstone modes at two orthogonal wave vectors ($\theta=\piup/2$; $\phi=0$ or $\piup/2$) corresponding to the would-be Bragg peak positions of the suppressed magnetic domains (open dots). The latter are characterized by a small spin gap $\Delta_{\rm PG}$ that can be only marginally resolved in our TOF data. Along all intermediate directions, the spin gap is larger, reaching a maximum along the Brillouin-zone diagonals.

To measure the angular dependence of the spin gap directly, we performed additional measurements at the \textsc{Panda} spectrometer within the $\theta=\pi/2$ plane along several radial directions relative to the zone center for a number of different $\phi$ angles. The measurement at each angle was done in three steps. First, we performed an energy scan at $|\mathbf{Q}|=q_\text{h}$ (assuming that the locus of the dispersion minima is spherical) to measure the approximate onset energy of magnetic scattering. Then, we did a radial scan at an energy characterized by the maximal slope of the spectrum, in order to find the true location of the dispersion minimum that could slightly deviate from $q_\text{h}$ for various angles. Finally, we repeated the energy scan at the new wave vector $|\mathbf{Q}|$ for every angle. These final energy scans are shown in Fig.~\ref{Fig:ExperimentGap}\,(a). It can be seen that the signal at $\phi=0$, at the location of the pseudo-Goldstone mode, has a finite energy that is clearly resolved from the elastic line. The maximum of inelastic scattering intensity is observed around 0.15\,meV. A more robust estimate of the gap energy results from fitting both the IN5 and \textsc{Panda} data from different radial directions, which resulted in the angular dependence of the spin-gap energy shown in Fig.\,\ref{Fig:ExperimentGap}\,(b). We can see that the results of both TOF and TAS measurements are in good agreement. The accuracy of our spin-gap measurement benefits from the extrapolation of the data taken away from the ordering vector, where the peak is better resolved, towards the low-energy region. We fitted the combined dataset in Fig.\,\ref{Fig:ExperimentGap}\,(b) with the simple empirical function $\Delta(\theta,\phi)=\sqrt{\Delta_{\rm PG}^2+A^2\sin^22\phi+B^2\sin^22\theta}$, which resulted in the experimental spin-gap value $\Delta_{\rm PG}=0.166(7)$\,meV. On the other hand, recent high-resolution powder data on ZnCr$_2$Se$_4$ demonstrate a gapless spin-wave spectrum down to at least 0.05\,meV \cite{ZajdelLi17}, which indicates that the anisotropy gap at the ordering vector $(0\,0\,q_\text{h})$ must be negligibly small, at least a factor of 3 smaller than $\Delta_{\rm PG}$.

\vspace{-3pt}\section{\hspace{-1ex}Discussion and conclusions}\label{Sec:Exp_details}

A remarkable hallmark of the observed pseudo-Goldstone modes is that LSWT predicts them to be gapless in the absence of magneto-crystalline anisotropy \cite{Kataoka87} at wave vectors where no magnetic Bragg reflections are found below $T_{\rm N}$. This is in contrast to magnetic soft modes observed, for example, in $\alpha$-CaCr$_2$O$_4$~\cite{TothLake12}, which originate from a proximity to another phase with a different magnetic ordering vector and therefore have a much larger gap that is well captured by LSWT. The existence of gapless modes without an underlying Bragg reflection would violate the Goldstone theorem, yet this contradiction can be resolved by taking into account quantum fluctuation corrections to the linear spectrum.

As long as the low-energy magnon spectrum is concerned, spin systems can be analyzed in the frame of the non-linear sigma model \cite{MilsteinSushkov11} or mapped onto a one-sublattice Heisenberg model. Therefore, to obtain an estimate of the pseudo-Goldstone spin gap, it is sufficient to consider a simplified model involving $J_1<0$, $J_2>0$, and $J_4>0$ interactions on a simple-cubic lattice,
\begin{eqnarray}
H = J_1 \sum_{\langle i,j\rangle} \mathbf{S}_{i} \mathbf{S}_{j} + J_2\!\!\sum_{\langle\!\langle i,j\rangle\!\rangle}\!\mathbf{S}_{i} \mathbf{S}_{j}+J_4\hspace{-5pt} \sum_{\langle\!\langle\!\langle\!\langle i,j \rangle\!\rangle\!\rangle\!\rangle}\hspace{-5pt}\mathbf{S}_{i} \mathbf{S}_{j},
\end{eqnarray}
where the summations are taken over the 1$^{\rm st}$, 2$^{\rm nd}$, and 4$^{\rm th}$ nearest neighbors, respectively. This minimal model results in the correct pitch angle $\gamma\approx0.23\piup$ and similar spin-wave velocities as those realized in ZnCr$_2$Se$_4$ for the following values of parameters:
$J_2=4$~meV, $J_4=1.8$~meV, and $J_1 = -4(J_2+J_4\cos\gamma)$.

\begin{figure}[b]\vspace{-5pt}
\begin{center}
\includegraphics[width=0.9\linewidth]{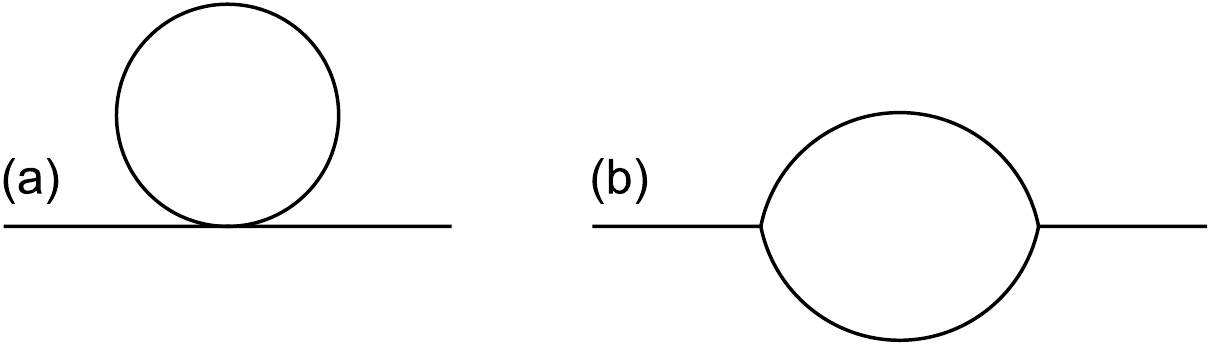}\vspace{-1em}
\end{center}
\caption{~The diagrams which give $1/S$ corrections to the spectrum: (a) $\omega^{(4)}_\mathbf{q}$ and (b) $\Sigma_{11}(\omega,\mathbf{q})$.\vspace{-6pt}}
\label{Fig:FeynmanDiags}
\end{figure}

To consider the magnon dispersion in the helical spin structure it is worthwhile to introduce local spin quantization axes $x$, $y$, $z$, such that $z\!\!\parallel\!\!\langle\mathbf{S}\rangle$ and $y\!\!\parallel\!\!\mathbf{Q}$. The Holstein-Primakoff transformation in the lowest $1/S$ order followed by the Bogoliubov transformation yields the magnon Hamiltonian  $H_{\rm m} = \sum_\mathbf{q}\omega_\mathbf{q}\alpha_\mathbf{q}^\dagger \alpha^{\phantom{\dagger}}_\mathbf{q}$ with the dispersion $\omega_\mathbf{q}=S\sqrt{(J_\mathbf{q}-J_\mathbf{Q})(J_\mathbf{q+Q}+J_\mathbf{q-Q}-2J_\mathbf{Q})/2}$, where $J_\mathbf{q}=\sum_{\mathbf R} J_\mathbf{R}\exp({\rm i}\mathbf{q}\cdot\mathbf{R})$. This magnon dispersion acquires six zeros at the $(0\,0\,\pm\!q_\text{h})$, $(0\,\pm\!q_\text{h}\,0)$, and $(\pm q_\text{h}\,0\,0)$ points. Quantum fluctuations given by the next orders in the $1/S$ expansion of the Holstein-Primakoff transformation open a gap at the $(0\,\pm\!q_\text{h}\,0)$ and $(\pm q_\text{h}\,0\,0)$ points orthogonal to the propagation vector \cite{ChubukovSachdev94, ChernyshevZhitomirsky06, ChernyshevZhitomirsky09, MilsteinSushkov11}. The lowest order of terms which gives such corrections is represented by three- and four-magnon processes shown in Fig.~\ref{Fig:FeynmanDiags}. Both $\omega^{(4)}_\mathbf{q}$ and ${\rm Re} \Sigma_{11}(\omega,\mathbf{q})$ terms contribute to the magnon dispersion. The imaginary part of the self-energy term ${\rm Im}\Sigma_{11}(\omega,\mathbf{q})$ describes the magnon damping $\Gamma_\mathbf{q}$:\vspace{-3pt}
\begin{equation}
\tilde\omega_{\mathbf{q}} + \mathrm{i}\Gamma_\mathbf{q} = \omega_\mathbf{q} + \omega^{(4)}_\mathbf{q} + \Sigma_{11}(\omega,\mathbf{q}).
\end{equation}

We calculated the magnon dispersion $\tilde \omega_\mathbf{q}$ at the helix propagation vector $\mathbf{Q}=(0\,0\,q_\text{h})$ and at the orthogonal wave vector $\mathbf{Q}'=(q_\text{h}\,0\,0)$. Our calculation shows that in the first case the quantum fluctuation corrections cancel out as expected in the $1/S$ expansion theory, whereas at the second vector $\mathbf{Q}'$ a spin-wave gap opens up with a magnitude $\Delta_{\rm PG} = 0.18$~meV, leading to a pseudo-Goldstone magnon observed in the experiment. The good agreement between the calculated and measured values of the spin gap suggests that the described quantum-fluctuation corrections to the LSWT are fully sufficient to describe the experimentally observed magnon spectrum. It is worth noting that the experimentally measured spin gap is expected to have an additional contribution dictated by the anisotropy energy, which is nonzero in any real material. However, the quantitative agreement of the calculated and measured gap magnitudes indicates that this contribution in ZnCr$_2$Se$_4$ is not dominant. Inclusion of a finite single-ion anisotropy term in our model would reduce the estimate of the spin gap from pure quantum-fluctuation effects even further, hence the experimental value of $\Delta_{\rm PG}$ should be seen as an upper estimate for the pseudo-Goldstone spin gap in a purely Heisenberg system.

It is interesting to discuss the implications of our results for collinear antiferromagnets. Stripe-like magnetic order with a $(\piup, 0)$ ordering vector, like the one realized in iron pnictides \cite{Dai15, Inosov16}, can be seen as a limiting case of a spin spiral with a pitch length of two lattice constants. In a simple frustrated $J_1$--$J_2$ Heisenberg model on a 2D square lattice, a collinear ground state is selected by the order-by-disorder mechanism \cite{SinghZheng03}. The linear spin-wave spectrum exhibits gapless modes not only at the $(\piup,0)$ ordering vector, but also at the $(0,\piup)$ and $(\piup,\piup)$ points. These accidental zero-energy modes develop finite spin gaps due to quantum fluctuations and are in this sense analogous to the pseudo-Goldstone modes discussed in our present work. Similarly, quantum-fluctuation effects were also claimed responsible for the reduction of the ordered moment in iron pnictides and significant modifications of their spin-wave spectrum \cite{UhrigHolt09}. However, there such changes appear to be so dramatic that they lead to a complete destruction of the pseudo-Goldstone modes. Thus, a recent experimental study performed on a uniaxially strained single crystal of BaFe$_2$As$_2$ clearly showed the absence of any measurable magnon intensity at the $(0,\piup)$ wave vector up to at least 40~meV, where the magnetic Bragg reflection was fully suppressed by detwinning \cite{LuPark14}. It therefore appears that the noncollinear helimagnetic ground state of ZnCr$_2$Se$_4$ and the large value of the spin ($S = 3/2$) that reduces quantum fluctuations are necessary prerequisites for observing clear pseudo-Goldstone modes with a small spin gap.

In summary, we have analyzed the phase diagram of the pyrochlore magnetic lattice for various coupling constants. The obtained theoretical phase diagram is in good agreement with the ground states of several $B$-site magnetic spinels that are known from experiments. Using inelastic neutron scattering, we investigated the complete spin-wave spectrum of ZnCr$_2$Se$_4$ in the whole of reciprocal space and across two orders of magnitude in energy. We found two distinct types of low-energy magnon modes: the Goldstone mode along the helix propagation vector at $(0\,0\,\pm\!q_\text{h})$ and two pseudo-Goldstone modes at $(\pm q_\text{h}\,0\,0)$ and $(0\,\pm\!q_\text{h}\,0)$. Our simulation in the framework of the LSWT provides a perfect description of the spectrum except at the pseudo-Goldstone points, where LSWT predicts a gapless mode, while the experiment shows a gap of the order $\Delta_{\rm PG} = 0.17$\,meV. This gap can be quantitatively reproduced in the calculations by including leading-order corrections due to quantum fluctuations arising from magnon-magnon scattering processes.

The scattering channels contributing to the magnon lifetime in insulators are usually very weak and require elaborate time-consuming experiments with a micro-electronvolt energy resolution to be quantified in direct measurements \cite{BayrakciKeller06}. At the same time, here we demonstrated that the pseudo-Goldstone magnon gap in helimagnets, first measured in our work, represents a well-defined and experimentally accessible quantity of the spin-wave spectrum that carries indirect information about such magnon-magnon interactions. In spite of its relatively simple experimental determination, the spin-gap magnitude $\Delta_{\rm PG}$ cannot be expressed analytically as a function of the Hamiltonian parameters and is therefore useful for testing spin-dynamical models beyond LSWT.

The existence of nearly gapless spin-wave modes propagating in the direction orthogonal to the ordering vector is expected to leave measurable signatures in the low-temperature thermodynamic and transport properties of the material, as well as in its response to local probes, at temperatures of the order of $\Delta_{\rm PG}\ll T_{\rm N}$. In a cubic system that has two pseudo-Goldstone modes per one Goldstone mode, the contribution of the former would dominate in all processes governed by low-energy magnon scattering, including but not limited to magnetic contributions to the low-temperature specific heat and magnon heat conduction~\cite{PrasaiTrump17}. In the presence of weak anisotropy, a two-gap structure in the magnon density of states is expected. To the best of our knowledge, these effects still await their detection in future experiments performed on samples in the single-domain magnetic state. Our present results should apply not only to simple helimagnets, but also to a much broader class of materials in which the magnetic propagation vector is spontaneously chosen from multiple structurally equivalent alternatives upon crossing a transition to the magnetically ordered state.

\vspace{-5pt}\section*{Acknowledgments}\vspace{-5pt}

We acknowledge fruitful discussions with A.~Yaresko and G.~Jackeli. This project was funded by the German Research Foundation (DFG) through the Collaborative Research Center SFB 1143 in Dresden [projects C03 (D.\,S.\,I.) and A06 (S.~R.)], individual research grant IN\,\mbox{209/4-1}, and the Trans\-regional Collaborative Research Center TRR~80 (Augsburg, Munich, Stuttgart). T.~M. and R.~T. \mbox{acknowledge} support from DFG-SFB 1170 ToCoTronics (project B04) and the ERC Starting Grant ERC-StG-Thomale-336012 ``Topolectrics''. The experiments at Oak Ridge National Laboratory's Spallation Neutron Source were sponsored by the Division of Scientific User Facilities, US DOE Office of Basic Energy Sciences.

Y.~V.~T. and Y.~A.\,O. contributed equally to this work. They performed neutron-scattering experiments together with A.~S.~C., P.~Y.~P., and D.~S.~I. and analyzed the data; T.~M., R.~T., S.~R., and D.\,V.~E. contributed to the theoretical interpretation of the results; D.~L.~A., J.~O., A. S., and A.~P. provided instrument support at the neutron sources; V.~T., V.~F., and A.~L. synthesized the samples; Y.~V.~T., S.~R., D.\,V.~E., and D.~S.~I. wrote the paper with contributions from all coauthors.

\bibliography{ZnCr2Se4}\vspace{-2pt}

\begin{thebibliography}{55}%
\makeatletter
\providecommand \@ifxundefined [1]{%
 \@ifx{#1\undefined}
}%
\providecommand \@ifnum [1]{%
 \ifnum #1\expandafter \@firstoftwo
 \else \expandafter \@secondoftwo
 \fi
}%
\providecommand \@ifx [1]{%
 \ifx #1\expandafter \@firstoftwo
 \else \expandafter \@secondoftwo
 \fi
}%
\providecommand \natexlab [1]{#1}%
\providecommand \enquote  [1]{``#1''}%
\providecommand \bibnamefont  [1]{#1}%
\providecommand \bibfnamefont [1]{#1}%
\providecommand \citenamefont [1]{#1}%
\providecommand \href@noop [0]{\@secondoftwo}%
\providecommand \href [0]{\begingroup \@sanitize@url \@href}%
\providecommand \@href[1]{\@@startlink{#1}\@@href}%
\providecommand \@@href[1]{\endgroup#1\@@endlink}%
\providecommand \@sanitize@url [0]{\catcode `\\12\catcode `\$12\catcode
  `\&12\catcode `\#12\catcode `\^12\catcode `\_12\catcode `\%12\relax}%
\providecommand \@@startlink[1]{}%
\providecommand \@@endlink[0]{}%
\providecommand \url  [0]{\begingroup\@sanitize@url \@url }%
\providecommand \@url [1]{\endgroup\@href {#1}{\urlprefix }}%
\providecommand \urlprefix  [0]{URL }%
\providecommand \Eprint [0]{\href }%
\providecommand \doibase [0]{http://dx.doi.org/}%
\providecommand \selectlanguage [0]{\@gobble}%
\providecommand \bibinfo  [0]{\@secondoftwo}%
\providecommand \bibfield  [0]{\@secondoftwo}%
\providecommand \translation [1]{[#1]}%
\providecommand \BibitemOpen [0]{}%
\providecommand \bibitemStop [0]{}%
\providecommand \bibitemNoStop [0]{.\EOS\space}%
\providecommand \EOS [0]{\spacefactor3000\relax}%
\providecommand \BibitemShut  [1]{\csname bibitem#1\endcsname}%
\let\auto@bib@innerbib\@empty
%</preamble>
\bibitem [{\citenamefont {Moessner}\ and\ \citenamefont
  {Chalker}(1998)}]{MoessnerChalker98}%
  \BibitemOpen
  \bibfield  {author} {\bibinfo {author} {\bibfnamefont {R.}~\bibnamefont
  {Moessner}}\ and\ \bibinfo {author} {\bibfnamefont {J.~T.}\ \bibnamefont
  {Chalker}},\ }\bibfield  {title} {\enquote {\bibinfo {title}
  {\textit{Properties of a classical spin liquid: The Heisenberg pyrochlore
  antiferromagnet}},}\ }\href {\doibase 10.1103/PhysRevLett.80.2929} {\bibfield
   {journal} {\bibinfo  {journal} {Phys. Rev. Lett.}\ }\textbf {\bibinfo
  {volume} {80}},\ \bibinfo {pages} {2929--2932} (\bibinfo {year}
  {1998})}\BibitemShut {NoStop}%
\bibitem [{\citenamefont {Reimers}\ \emph {et~al.}(1991)\citenamefont
  {Reimers}, \citenamefont {Berlinsky},\ and\ \citenamefont
  {Shi}}]{ReimersBerlinsky91}%
  \BibitemOpen
  \bibfield  {author} {\bibinfo {author} {\bibfnamefont {J.~N.}\ \bibnamefont
  {Reimers}}, \bibinfo {author} {\bibfnamefont {A.~J.}\ \bibnamefont
  {Berlinsky}}, \ and\ \bibinfo {author} {\bibfnamefont {A.-C.}\ \bibnamefont
  {Shi}},\ }\bibfield  {title} {\enquote {\bibinfo {title} {\textit{Mean-field
  approach to magnetic ordering in highly frustrated pyrochlores}},}\ }\href
  {\doibase 10.1103/PhysRevB.43.865} {\bibfield  {journal} {\bibinfo  {journal}
  {Phys. Rev.~B}\ }\textbf {\bibinfo {volume} {43}},\ \bibinfo {pages}
  {865--878} (\bibinfo {year} {1991})}\BibitemShut {NoStop}%
\bibitem [{\citenamefont {Lapa}\ and\ \citenamefont {Henley}()}]{LapaHenley12}%
  \BibitemOpen
  \bibfield  {author} {\bibinfo {author} {\bibfnamefont {M.~F.}\ \bibnamefont
  {Lapa}}\ and\ \bibinfo {author} {\bibfnamefont {C.~L.}\ \bibnamefont
  {Henley}},\ }\href@noop {} {\enquote {\bibinfo {title} {\textit{Ground states
  of the classical antiferromagnet on the pyrochlore lattice}},}\ }\bibinfo
  {howpublished} {\href{http://arxiv.org/abs/arXiv:1210.6810}{arXiv:1210.6810}
  (unpublished).}\BibitemShut {Stop}%
\bibitem [{\citenamefont {Takata}\ \emph {et~al.}()\citenamefont {Takata},
  \citenamefont {Momoi},\ and\ \citenamefont {Oshikawa}}]{TakataMomoi15}%
  \BibitemOpen
  \bibfield  {author} {\bibinfo {author} {\bibfnamefont {E.}~\bibnamefont
  {Takata}}, \bibinfo {author} {\bibfnamefont {T.}~\bibnamefont {Momoi}}, \
  and\ \bibinfo {author} {\bibfnamefont {M.}~\bibnamefont {Oshikawa}},\
  }\href@noop {} {\enquote {\bibinfo {title} {\textit{Nematic ordering in
  pyrochlore antiferromagnets: high-field phase of chromium spinel oxides}},}\
  }\bibinfo {howpublished}
  {\href{http://arxiv.org/abs/arXiv:1510.02373}{arXiv:1510.02373}
  (unpublished).}\BibitemShut {Stop}%
\bibitem [{\citenamefont {Okubo}\ \emph {et~al.}(2011)\citenamefont {Okubo},
  \citenamefont {Nguyen},\ and\ \citenamefont {Kawamura}}]{OkuboNguyen11}%
  \BibitemOpen
  \bibfield  {author} {\bibinfo {author} {\bibfnamefont {T.}~\bibnamefont
  {Okubo}}, \bibinfo {author} {\bibfnamefont {T.~H.}\ \bibnamefont {Nguyen}}, \
  and\ \bibinfo {author} {\bibfnamefont {H.}~\bibnamefont {Kawamura}},\
  }\bibfield  {title} {\enquote {\bibinfo {title} {\textit{Cubic and noncubic
  multiple-$q$ states in the Heisenberg antiferromagnet on the pyrochlore
  lattice}},}\ }\href {\doibase 10.1103/PhysRevB.84.144432} {\bibfield
  {journal} {\bibinfo  {journal} {Phys. Rev.~B}\ }\textbf {\bibinfo {volume}
  {84}},\ \bibinfo {pages} {144432} (\bibinfo {year} {2011})}\BibitemShut
  {NoStop}%
\bibitem [{\citenamefont {Conlon}\ and\ \citenamefont
  {Chalker}(2010)}]{ConlonChalker10}%
  \BibitemOpen
  \bibfield  {author} {\bibinfo {author} {\bibfnamefont {P.~H.}\ \bibnamefont
  {Conlon}}\ and\ \bibinfo {author} {\bibfnamefont {J.~T.}\ \bibnamefont
  {Chalker}},\ }\bibfield  {title} {\enquote {\bibinfo {title} {\textit{Absent
  pinch points and emergent clusters: Further neighbor interactions in the
  pyrochlore Heisenberg antiferromagnet}},}\ }\href {\doibase
  10.1103/PhysRevB.81.224413} {\bibfield  {journal} {\bibinfo  {journal} {Phys.
  Rev.~B}\ }\textbf {\bibinfo {volume} {81}},\ \bibinfo {pages} {224413}
  (\bibinfo {year} {2010})}\BibitemShut {NoStop}%
\bibitem [{\citenamefont {Yaresko}(2008)}]{Yaresko08}%
  \BibitemOpen
  \bibfield  {author} {\bibinfo {author} {\bibfnamefont {A.~N.}\ \bibnamefont
  {Yaresko}},\ }\bibfield  {title} {\enquote {\bibinfo {title}
  {\textit{Electronic band structure and exchange coupling constants in
  $A{\kern.5pt}\text{Cr}_\text{\!2}X_\text{4}$ spinels ($A$\,\,=~Zn, Cd, Hg;
  $X$\,=~O, S, Se)}},}\ }\href {\doibase 10.1103/PhysRevB.77.115106} {\bibfield
   {journal} {\bibinfo  {journal} {Phys. Rev.~B}\ }\textbf {\bibinfo {volume}
  {77}},\ \bibinfo {pages} {115106} (\bibinfo {year} {2008})}\BibitemShut
  {NoStop}%
\bibitem [{\citenamefont {Siratori}(1971)}]{Siratori71}%
  \BibitemOpen
  \bibfield  {author} {\bibinfo {author} {\bibfnamefont {K.}~\bibnamefont
  {Siratori}},\ }\bibfield  {title} {\enquote {\bibinfo {title}
  {\textit{Magnetic resonance of ZnCr$_\text{\!2}$Se$_\text{4}$ with screw spin
  structure}},}\ }\href {\doibase 10.1143/JPSJ.30.709} {\bibfield  {journal}
  {\bibinfo  {journal} {J.~Phys. Soc. Jpn.}\ }\textbf {\bibinfo {volume}
  {30}},\ \bibinfo {pages} {709--719} (\bibinfo {year} {1971})}\BibitemShut
  {NoStop}%
\bibitem [{\citenamefont {Zhang}\ \emph {et~al.}(2016)\citenamefont {Zhang},
  \citenamefont {Su}, \citenamefont {Chen}, \citenamefont {Zhang},
  \citenamefont {Mei}, \citenamefont {Yang}, \citenamefont {Dai},\ and\
  \citenamefont {Pi}}]{ZhangSu16}%
  \BibitemOpen
  \bibfield  {author} {\bibinfo {author} {\bibfnamefont {P.}~\bibnamefont
  {Zhang}}, \bibinfo {author} {\bibfnamefont {F.}~\bibnamefont {Su}}, \bibinfo
  {author} {\bibfnamefont {X.}~\bibnamefont {Chen}}, \bibinfo {author}
  {\bibfnamefont {S.}~\bibnamefont {Zhang}}, \bibinfo {author} {\bibfnamefont
  {H.}~\bibnamefont {Mei}}, \bibinfo {author} {\bibfnamefont {Z.}~\bibnamefont
  {Yang}}, \bibinfo {author} {\bibfnamefont {J.}~\bibnamefont {Dai}}, \ and\
  \bibinfo {author} {\bibfnamefont {L.}~\bibnamefont {Pi}},\ }\bibfield
  {title} {\enquote {\bibinfo {title} {\textit{Observation of magnetic
  resonance and Faraday rotation in a spinel ZnCr$_\text{\!2}$Se$_\text{4}$
  single crystal in the terahertz region}},}\ }\href {\doibase
  10.7567/APEX.9.102401} {\bibfield  {journal} {\bibinfo  {journal} {Appl.
  Phys. Express}\ }\textbf {\bibinfo {volume} {9}},\ \bibinfo {pages} {102401}
  (\bibinfo {year} {2016})}\BibitemShut {NoStop}%
\bibitem [{\citenamefont {Aoyama}\ and\ \citenamefont
  {Kawamura}(2016)}]{AoyamaKawamura16}%
  \BibitemOpen
  \bibfield  {author} {\bibinfo {author} {\bibfnamefont {K.}~\bibnamefont
  {Aoyama}}\ and\ \bibinfo {author} {\bibfnamefont {H.}~\bibnamefont
  {Kawamura}},\ }\bibfield  {title} {\enquote {\bibinfo {title}
  {\textit{Spin-lattice-coupled order in Heisenberg antiferromagnets on the
  pyrochlore lattice}},}\ }\href {\doibase 10.1103/PhysRevLett.116.257201}
  {\bibfield  {journal} {\bibinfo  {journal} {Phys. Rev. Lett.}\ }\textbf
  {\bibinfo {volume} {116}},\ \bibinfo {pages} {257201} (\bibinfo {year}
  {2016})}\BibitemShut {NoStop}%
\bibitem [{\citenamefont {Clark}\ \emph {et~al.}(2014)\citenamefont {Clark},
  \citenamefont {Nilsen}, \citenamefont {Kermarrec}, \citenamefont {Ehlers},
  \citenamefont {Knight}, \citenamefont {Harrison}, \citenamefont {Attfield},\
  and\ \citenamefont {Gaulin}}]{ClarkNilsen14}%
  \BibitemOpen
  \bibfield  {author} {\bibinfo {author} {\bibfnamefont {L.}~\bibnamefont
  {Clark}}, \bibinfo {author} {\bibfnamefont {G.~J.}\ \bibnamefont {Nilsen}},
  \bibinfo {author} {\bibfnamefont {E.}~\bibnamefont {Kermarrec}}, \bibinfo
  {author} {\bibfnamefont {G.}~\bibnamefont {Ehlers}}, \bibinfo {author}
  {\bibfnamefont {K.~S.}\ \bibnamefont {Knight}}, \bibinfo {author}
  {\bibfnamefont {A.}~\bibnamefont {Harrison}}, \bibinfo {author}
  {\bibfnamefont {J.~P.}\ \bibnamefont {Attfield}}, \ and\ \bibinfo {author}
  {\bibfnamefont {B.~D.}\ \bibnamefont {Gaulin}},\ }\bibfield  {title}
  {\enquote {\bibinfo {title} {\textit{From spin glass to quantum spin liquid
  ground states in molybdate pyrochlores}},}\ }\href {\doibase
  10.1103/PhysRevLett.113.117201} {\bibfield  {journal} {\bibinfo  {journal}
  {Phys. Rev. Lett.}\ }\textbf {\bibinfo {volume} {113}},\ \bibinfo {pages}
  {117201} (\bibinfo {year} {2014})}\BibitemShut {NoStop}%
\bibitem [{\citenamefont {Iqbal}\ \emph {et~al.}()\citenamefont {Iqbal},
  \citenamefont {M\"uller}, \citenamefont {Riedl}, \citenamefont {Reuther},
  \citenamefont {Rachel}, \citenamefont {Valenti}, \citenamefont {Gingras},
  \citenamefont {Thomale},\ and\ \citenamefont {Jeschke}}]{IqbalMueller17}%
  \BibitemOpen
  \bibfield  {author} {\bibinfo {author} {\bibfnamefont {Y.}~\bibnamefont
  {Iqbal}}, \bibinfo {author} {\bibfnamefont {T.}~\bibnamefont {M\"uller}},
  \bibinfo {author} {\bibfnamefont {K.}~\bibnamefont {Riedl}}, \bibinfo
  {author} {\bibfnamefont {J.}~\bibnamefont {Reuther}}, \bibinfo {author}
  {\bibfnamefont {S.}~\bibnamefont {Rachel}}, \bibinfo {author} {\bibfnamefont
  {R.}~\bibnamefont {Valenti}}, \bibinfo {author} {\bibfnamefont {M.~J.~P.}\
  \bibnamefont {Gingras}}, \bibinfo {author} {\bibfnamefont {R.}~\bibnamefont
  {Thomale}}, \ and\ \bibinfo {author} {\bibfnamefont {H.~O.}\ \bibnamefont
  {Jeschke}},\ }\href@noop {} {\enquote {\bibinfo {title} {\textit{Signatures
  of a gearwheel quantum spin liquid in a spin-$1{\kern-.5pt}/{\kern-.5pt}2$
  pyrochlore molybdate Heisenberg antiferromagnet}},}\ }\bibinfo {howpublished}
  {\href{http://arxiv.org/abs/arXiv:1705.05291}{arXiv:1705.05291}
  (unpublished).}\BibitemShut {Stop}%
\bibitem [{\citenamefont {Matsuda}\ \emph {et~al.}(2007)\citenamefont
  {Matsuda}, \citenamefont {Ueda}, \citenamefont {Kikkawa}, \citenamefont
  {Tanaka}, \citenamefont {Katsumata}, \citenamefont {Narumi}, \citenamefont
  {Inami}, \citenamefont {Ueda},\ and\ \citenamefont {Lee}}]{MatsudaUeda07}%
  \BibitemOpen
  \bibfield  {author} {\bibinfo {author} {\bibfnamefont {M.}~\bibnamefont
  {Matsuda}}, \bibinfo {author} {\bibfnamefont {H.}~\bibnamefont {Ueda}},
  \bibinfo {author} {\bibfnamefont {A.}~\bibnamefont {Kikkawa}}, \bibinfo
  {author} {\bibfnamefont {Y.}~\bibnamefont {Tanaka}}, \bibinfo {author}
  {\bibfnamefont {K.}~\bibnamefont {Katsumata}}, \bibinfo {author}
  {\bibfnamefont {Y.}~\bibnamefont {Narumi}}, \bibinfo {author} {\bibfnamefont
  {T.}~\bibnamefont {Inami}}, \bibinfo {author} {\bibfnamefont
  {Y.}~\bibnamefont {Ueda}}, \ and\ \bibinfo {author} {\bibfnamefont {S.-H.}\
  \bibnamefont {Lee}},\ }\bibfield  {title} {\enquote {\bibinfo {title}
  {\textit{Spin-lattice instability to a fractional magnetization state in the
  spinel HgCr$_\text{\!2}$O$_\text{4}$}},}\ }\href {\doibase 10.1038/nphys586}
  {\bibfield  {journal} {\bibinfo  {journal} {Nat. Phys.}\ }\textbf {\bibinfo
  {volume} {3}},\ \bibinfo {pages} {397} (\bibinfo {year} {2007})}\BibitemShut
  {NoStop}%
\bibitem [{\citenamefont {Yokaichiya}\ \emph {et~al.}(2009)\citenamefont
  {Yokaichiya}, \citenamefont {Krimmel}, \citenamefont {Tsurkan}, \citenamefont
  {Margiolaki}, \citenamefont {Thompson}, \citenamefont {Bordallo},
  \citenamefont {Buchsteiner}, \citenamefont {St\"u\ss{}er}, \citenamefont
  {Argyriou},\ and\ \citenamefont {Loidl}}]{YokaichiyaKrimmel09}%
  \BibitemOpen
  \bibfield  {author} {\bibinfo {author} {\bibfnamefont {F.}~\bibnamefont
  {Yokaichiya}}, \bibinfo {author} {\bibfnamefont {A.}~\bibnamefont {Krimmel}},
  \bibinfo {author} {\bibfnamefont {V.}~\bibnamefont {Tsurkan}}, \bibinfo
  {author} {\bibfnamefont {I.}~\bibnamefont {Margiolaki}}, \bibinfo {author}
  {\bibfnamefont {P.}~\bibnamefont {Thompson}}, \bibinfo {author}
  {\bibfnamefont {H.~N.}\ \bibnamefont {Bordallo}}, \bibinfo {author}
  {\bibfnamefont {A.}~\bibnamefont {Buchsteiner}}, \bibinfo {author}
  {\bibfnamefont {N.}~\bibnamefont {St\"u\ss{}er}}, \bibinfo {author}
  {\bibfnamefont {D.~N.}\ \bibnamefont {Argyriou}}, \ and\ \bibinfo {author}
  {\bibfnamefont {A.}~\bibnamefont {Loidl}},\ }\bibfield  {title} {\enquote
  {\bibinfo {title} {\textit{Spin-driven phase transitions in
  ZnCr$_\text{\!2}$Se$_\text{4}$ and ZnCr$_\text{\!2}$S$_\text{4}$ probed by
  high-resolution synchrotron x-ray and neutron powder diffraction}},}\ }\href
  {\doibase 10.1103/PhysRevB.79.064423} {\bibfield  {journal} {\bibinfo
  {journal} {Phys. Rev.~B}\ }\textbf {\bibinfo {volume} {79}},\ \bibinfo
  {pages} {064423} (\bibinfo {year} {2009})}\BibitemShut {NoStop}%
\bibitem [{\citenamefont {Menyuk}\ \emph {et~al.}(1966)\citenamefont {Menyuk},
  \citenamefont {Dwight}, \citenamefont {Arnott},\ and\ \citenamefont
  {Wold}}]{MenyukDwight66}%
  \BibitemOpen
  \bibfield  {author} {\bibinfo {author} {\bibfnamefont {N.}~\bibnamefont
  {Menyuk}}, \bibinfo {author} {\bibfnamefont {K.}~\bibnamefont {Dwight}},
  \bibinfo {author} {\bibfnamefont {R.~J.}\ \bibnamefont {Arnott}}, \ and\
  \bibinfo {author} {\bibfnamefont {A.}~\bibnamefont {Wold}},\ }\bibfield
  {title} {\enquote {\bibinfo {title} {\textit{Ferromagnetism in
  CdCr$_\text{\!2}$Se$_\text{4}$ and CdCr$_\text{\!2}$S$_\text{4}$}},}\ }\href
  {\doibase 10.1063/1.1708484} {\bibfield  {journal} {\bibinfo  {journal}
  {J.~Appl. Phys.}\ }\textbf {\bibinfo {volume} {37}},\ \bibinfo {pages} {1387}
  (\bibinfo {year} {1966})}\BibitemShut {NoStop}%
\bibitem [{\citenamefont {Hastings}\ and\ \citenamefont
  {Corliss}(1968)}]{HastingsCorliss68}%
  \BibitemOpen
  \bibfield  {author} {\bibinfo {author} {\bibfnamefont {J.~M.}\ \bibnamefont
  {Hastings}}\ and\ \bibinfo {author} {\bibfnamefont {L.~M.}\ \bibnamefont
  {Corliss}},\ }\bibfield  {title} {\enquote {\bibinfo {title}
  {\textit{Magnetic structure and metamagnetism of
  HgCr$_\text{\!2}$S$_\text{4}$}},}\ }\href {\doibase
  10.1016/0022-3697(68)90249-7} {\bibfield  {journal} {\bibinfo  {journal} {J.
  Phys. Chem. Solids}\ }\textbf {\bibinfo {volume} {29}},\ \bibinfo {pages}
  {9--14} (\bibinfo {year} {1968})}\BibitemShut {NoStop}%
\bibitem [{\citenamefont {Tsurkan}\ \emph {et~al.}(2006)\citenamefont
  {Tsurkan}, \citenamefont {Hemberger}, \citenamefont {Krimmel}, \citenamefont
  {Krug~von Nidda}, \citenamefont {Lunkenheimer}, \citenamefont {Weber},
  \citenamefont {Zestrea},\ and\ \citenamefont {Loidl}}]{TsurkanHemberger06}%
  \BibitemOpen
  \bibfield  {author} {\bibinfo {author} {\bibfnamefont {V.}~\bibnamefont
  {Tsurkan}}, \bibinfo {author} {\bibfnamefont {J.}~\bibnamefont {Hemberger}},
  \bibinfo {author} {\bibfnamefont {A.}~\bibnamefont {Krimmel}}, \bibinfo
  {author} {\bibfnamefont {H.-A.}\ \bibnamefont {Krug~von Nidda}}, \bibinfo
  {author} {\bibfnamefont {P.}~\bibnamefont {Lunkenheimer}}, \bibinfo {author}
  {\bibfnamefont {S.}~\bibnamefont {Weber}}, \bibinfo {author} {\bibfnamefont
  {V.}~\bibnamefont {Zestrea}}, \ and\ \bibinfo {author} {\bibfnamefont
  {A.}~\bibnamefont {Loidl}},\ }\bibfield  {title} {\enquote {\bibinfo {title}
  {\textit{Experimental evidence for competition between antiferromagnetic and
  ferromagnetic correlations in HgCr$_\text{\!2}$S$_\text{4}$}},}\ }\href
  {\doibase 10.1103/PhysRevB.73.224442} {\bibfield  {journal} {\bibinfo
  {journal} {Phys. Rev.~B}\ }\textbf {\bibinfo {volume} {73}},\ \bibinfo
  {pages} {224442} (\bibinfo {year} {2006})}\BibitemShut {NoStop}%
\bibitem [{\citenamefont {Plumier}(1966)}]{Plumier66}%
  \BibitemOpen
  \bibfield  {author} {\bibinfo {author} {\bibfnamefont {L.~J.}\ \bibnamefont
  {Plumier}},\ }\bibfield  {title} {\enquote {\bibinfo {title} {\textit{Neutron
  diffraction study of helimagnetic spinel ZnCr$_\text{\!2}$Se$_\text{4}$}},}\
  }\href {\doibase 10.1063/1.1708540} {\bibfield  {journal} {\bibinfo
  {journal} {J. Appl. Phys.}\ }\textbf {\bibinfo {volume} {37}},\ \bibinfo
  {pages} {964--965} (\bibinfo {year} {1966})}\BibitemShut {NoStop}%
\bibitem [{\citenamefont {Cameron}\ \emph {et~al.}(2016)\citenamefont
  {Cameron}, \citenamefont {Tymoshenko}, \citenamefont {Portnichenko},
  \citenamefont {Gavilano}, \citenamefont {Tsurkan}, \citenamefont {Felea},
  \citenamefont {Loidl}, \citenamefont {Zherlitsyn}, \citenamefont {Wosnitza},\
  and\ \citenamefont {Inosov}}]{CameronTymoshenko16}%
  \BibitemOpen
  \bibfield  {author} {\bibinfo {author} {\bibfnamefont {A.~S.}\ \bibnamefont
  {Cameron}}, \bibinfo {author} {\bibfnamefont {Y.~V.}\ \bibnamefont
  {Tymoshenko}}, \bibinfo {author} {\bibfnamefont {P.~Y.}\ \bibnamefont
  {Portnichenko}}, \bibinfo {author} {\bibfnamefont {J.}~\bibnamefont
  {Gavilano}}, \bibinfo {author} {\bibfnamefont {V.}~\bibnamefont {Tsurkan}},
  \bibinfo {author} {\bibfnamefont {V.}~\bibnamefont {Felea}}, \bibinfo
  {author} {\bibfnamefont {A.}~\bibnamefont {Loidl}}, \bibinfo {author}
  {\bibfnamefont {S.}~\bibnamefont {Zherlitsyn}}, \bibinfo {author}
  {\bibfnamefont {J.}~\bibnamefont {Wosnitza}}, \ and\ \bibinfo {author}
  {\bibfnamefont {D.~S.}\ \bibnamefont {Inosov}},\ }\bibfield  {title}
  {\enquote {\bibinfo {title} {\textit{Magnetic phase diagram of the
  helimagnetic spinel compound ZnCr$_\text{\!2}$Se$_\text{4}$ revisited by
  small-angle neutron scattering}},}\ }\href {\doibase
  10.1088/0953-8984/28/14/146001} {\bibfield  {journal} {\bibinfo  {journal}
  {J.~Phys.: Condens. Matter}\ }\textbf {\bibinfo {volume} {28}},\ \bibinfo
  {pages} {146001} (\bibinfo {year} {2016})}\BibitemShut {NoStop}%
\bibitem [{\citenamefont {Wojtowicz}(1969)}]{Wojtowicz69}%
  \BibitemOpen
  \bibfield  {author} {\bibinfo {author} {\bibfnamefont {P.}~\bibnamefont
  {Wojtowicz}},\ }\bibfield  {title} {\enquote {\bibinfo {title}
  {\textit{Semiconducting ferromagnetic spinels}},}\ }\href {\doibase
  10.1109/TMAG.1969.1066666} {\bibfield  {journal} {\bibinfo  {journal} {IEEE
  Trans. Magn.}\ }\textbf {\bibinfo {volume} {5}},\ \bibinfo {pages} {840}
  (\bibinfo {year} {1969})}\BibitemShut {NoStop}%
\bibitem [{\citenamefont {Ueda}\ \emph {et~al.}(2006)\citenamefont {Ueda},
  \citenamefont {Mitamura}, \citenamefont {Goto},\ and\ \citenamefont
  {Ueda}}]{UedaMitamura06}%
  \BibitemOpen
  \bibfield  {author} {\bibinfo {author} {\bibfnamefont {H.}~\bibnamefont
  {Ueda}}, \bibinfo {author} {\bibfnamefont {H.}~\bibnamefont {Mitamura}},
  \bibinfo {author} {\bibfnamefont {T.}~\bibnamefont {Goto}}, \ and\ \bibinfo
  {author} {\bibfnamefont {Y.}~\bibnamefont {Ueda}},\ }\bibfield  {title}
  {\enquote {\bibinfo {title} {\textit{Successive field-induced transitions in
  a frustrated antiferromagnet HgCr$_\text{\!2}$O$_\text{4}$}},}\ }\href
  {\doibase 10.1103/PhysRevB.73.094415} {\bibfield  {journal} {\bibinfo
  {journal} {Phys. Rev.~B}\ }\textbf {\bibinfo {volume} {73}},\ \bibinfo
  {pages} {094415} (\bibinfo {year} {2006})}\BibitemShut {NoStop}%
\bibitem [{\citenamefont {Akimitsu}\ \emph {et~al.}(1978)\citenamefont
  {Akimitsu}, \citenamefont {Siratori}, \citenamefont {Shirane}, \citenamefont
  {Iizumi},\ and\ \citenamefont {Watanabe}}]{AkimitsuSiratori78}%
  \BibitemOpen
  \bibfield  {author} {\bibinfo {author} {\bibfnamefont {J.}~\bibnamefont
  {Akimitsu}}, \bibinfo {author} {\bibfnamefont {K.}~\bibnamefont {Siratori}},
  \bibinfo {author} {\bibfnamefont {G.}~\bibnamefont {Shirane}}, \bibinfo
  {author} {\bibfnamefont {M.}~\bibnamefont {Iizumi}}, \ and\ \bibinfo {author}
  {\bibfnamefont {T.}~\bibnamefont {Watanabe}},\ }\bibfield  {title} {\enquote
  {\bibinfo {title} {\textit{Neutron scattering study of
  ZnCr$_\text{\!2}$Se$_\text{4}$ with screw spin structure}},}\ }\href
  {\doibase 10.1143/JPSJ.44.172} {\bibfield  {journal} {\bibinfo  {journal}
  {J.~Phys. Soc. Jpn.}\ }\textbf {\bibinfo {volume} {44}},\ \bibinfo {pages}
  {172--180} (\bibinfo {year} {1978})}\BibitemShut {NoStop}%
\bibitem [{\citenamefont {Yoshimori}(1959)}]{Yoshimori59}%
  \BibitemOpen
  \bibfield  {author} {\bibinfo {author} {\bibfnamefont {A.}~\bibnamefont
  {Yoshimori}},\ }\bibfield  {title} {\enquote {\bibinfo {title} {\textit{A new
  type of antiferromagnetic structure in the rutile type crystal}},}\ }\href
  {\doibase 10.1143/JPSJ.14.807} {\bibfield  {journal} {\bibinfo  {journal}
  {J.~Phys. Soc. Jpn.}\ }\textbf {\bibinfo {volume} {14}},\ \bibinfo {pages}
  {807--821} (\bibinfo {year} {1959})}\BibitemShut {NoStop}%
\bibitem [{\citenamefont {Togawa}\ \emph {et~al.}(2016)\citenamefont {Togawa},
  \citenamefont {Kousaka}, \citenamefont {Inoue},\ and\ \citenamefont
  {Kishine}}]{TogawaKousaka16}%
  \BibitemOpen
  \bibfield  {author} {\bibinfo {author} {\bibfnamefont {Y.}~\bibnamefont
  {Togawa}}, \bibinfo {author} {\bibfnamefont {Y.}~\bibnamefont {Kousaka}},
  \bibinfo {author} {\bibfnamefont {K.}~\bibnamefont {Inoue}}, \ and\ \bibinfo
  {author} {\bibfnamefont {J.-I.}\ \bibnamefont {Kishine}},\ }\bibfield
  {title} {\enquote {\bibinfo {title} {\textit{Symmetry, structure, and
  dynamics of monoaxial chiral magnets}},}\ }\href {\doibase
  10.7566/JPSJ.85.112001} {\bibfield  {journal} {\bibinfo  {journal} {J.~Phys.
  Soc. Jpn.}\ }\textbf {\bibinfo {volume} {85}},\ \bibinfo {pages} {112001}
  (\bibinfo {year} {2016})}\BibitemShut {NoStop}%
\bibitem [{\citenamefont {Tokura}\ and\ \citenamefont
  {Seki}(2010)}]{TokuraSeki10}%
  \BibitemOpen
  \bibfield  {author} {\bibinfo {author} {\bibfnamefont {Y.}~\bibnamefont
  {Tokura}}\ and\ \bibinfo {author} {\bibfnamefont {S.}~\bibnamefont {Seki}},\
  }\bibfield  {title} {\enquote {\bibinfo {title} {\textit{Multiferroics with
  spiral spin orders}},}\ }\href {\doibase 10.1002/adma.200901961} {\bibfield
  {journal} {\bibinfo  {journal} {Adv. Mater.}\ }\textbf {\bibinfo {volume}
  {22}},\ \bibinfo {pages} {1554--1565} (\bibinfo {year} {2010})}\BibitemShut
  {NoStop}%
\bibitem [{\citenamefont {Kimura}(2012)}]{Kimura12}%
  \BibitemOpen
  \bibfield  {author} {\bibinfo {author} {\bibfnamefont {Tsuyoshi}\
  \bibnamefont {Kimura}},\ }\bibfield  {title} {\enquote {\bibinfo {title}
  {\textit{Magnetoelectric hexaferrites}},}\ }\href {\doibase
  10.1146/annurev-conmatphys-020911-125101} {\bibfield  {journal} {\bibinfo
  {journal} {Annu. Rev. Condens. Matter Phys.}\ }\textbf {\bibinfo {volume}
  {3}},\ \bibinfo {pages} {93--110} (\bibinfo {year} {2012})}\BibitemShut
  {NoStop}%
\bibitem [{\citenamefont {Rudolf}\ \emph {et~al.}(2007)\citenamefont {Rudolf},
  \citenamefont {Kant}, \citenamefont {Mayr}, \citenamefont {Hemberger},
  \citenamefont {Tsurkan},\ and\ \citenamefont {Loidl}}]{RudolfKant07}%
  \BibitemOpen
  \bibfield  {author} {\bibinfo {author} {\bibfnamefont {T.}~\bibnamefont
  {Rudolf}}, \bibinfo {author} {\bibfnamefont {C.}~\bibnamefont {Kant}},
  \bibinfo {author} {\bibfnamefont {F.}~\bibnamefont {Mayr}}, \bibinfo {author}
  {\bibfnamefont {J.}~\bibnamefont {Hemberger}}, \bibinfo {author}
  {\bibfnamefont {V.}~\bibnamefont {Tsurkan}}, \ and\ \bibinfo {author}
  {\bibfnamefont {A.}~\bibnamefont {Loidl}},\ }\bibfield  {title} {\enquote
  {\bibinfo {title} {\textit{Spin-phonon coupling in antiferromagnetic chromium
  spinels}},}\ }\href {\doibase 10.1088/1367-2630/9/3/076} {\bibfield
  {journal} {\bibinfo  {journal} {New J.~Phys.}\ }\textbf {\bibinfo {volume}
  {9}},\ \bibinfo {pages} {76} (\bibinfo {year} {2007})}\BibitemShut {NoStop}%
\bibitem [{\citenamefont {Zajdel}\ \emph {et~al.}(2017)\citenamefont {Zajdel},
  \citenamefont {Li}, \citenamefont {van Beek}, \citenamefont {Lappas},
  \citenamefont {Ziolkowska}, \citenamefont {Jaskiewicz}, \citenamefont
  {Stock},\ and\ \citenamefont {Green}}]{ZajdelLi17}%
  \BibitemOpen
  \bibfield  {author} {\bibinfo {author} {\bibfnamefont {P.}~\bibnamefont
  {Zajdel}}, \bibinfo {author} {\bibfnamefont {W.-Y.}\ \bibnamefont {Li}},
  \bibinfo {author} {\bibfnamefont {W.}~\bibnamefont {van Beek}}, \bibinfo
  {author} {\bibfnamefont {A.}~\bibnamefont {Lappas}}, \bibinfo {author}
  {\bibfnamefont {A.}~\bibnamefont {Ziolkowska}}, \bibinfo {author}
  {\bibfnamefont {S.}~\bibnamefont {Jaskiewicz}}, \bibinfo {author}
  {\bibfnamefont {C.}~\bibnamefont {Stock}}, \ and\ \bibinfo {author}
  {\bibfnamefont {M.~A.}\ \bibnamefont {Green}},\ }\bibfield  {title} {\enquote
  {\bibinfo {title} {\textit{Structure and magnetism in the bond-frustrated
  spinel ZnCr$_\text{\!2}$Se$_\text{4}$}},}\ }\href {\doibase
  10.1103/PhysRevB.95.134401} {\bibfield  {journal} {\bibinfo  {journal} {Phys.
  Rev.~B}\ }\textbf {\bibinfo {volume} {95}},\ \bibinfo {pages} {134401}
  (\bibinfo {year} {2017})}\BibitemShut {NoStop}%
\bibitem [{\citenamefont {Abdul-Jabbar}\ \emph {et~al.}(2015)\citenamefont
  {Abdul-Jabbar}, \citenamefont {Sokolov}, \citenamefont {O'Neill},
  \citenamefont {Stock}, \citenamefont {Wermeille}, \citenamefont {Demmel},
  \citenamefont {Kr\"uger}, \citenamefont {Green}, \citenamefont
  {L\'evy-Bertrand}, \citenamefont {Grenier},\ and\ \citenamefont
  {Huxley}}]{AbdulJabbar15}%
  \BibitemOpen
  \bibfield  {author} {\bibinfo {author} {\bibfnamefont {G.}~\bibnamefont
  {Abdul-Jabbar}}, \bibinfo {author} {\bibfnamefont {D.~A.}\ \bibnamefont
  {Sokolov}}, \bibinfo {author} {\bibfnamefont {C.~D.}\ \bibnamefont
  {O'Neill}}, \bibinfo {author} {\bibfnamefont {C.}~\bibnamefont {Stock}},
  \bibinfo {author} {\bibfnamefont {D.}~\bibnamefont {Wermeille}}, \bibinfo
  {author} {\bibfnamefont {F.}~\bibnamefont {Demmel}}, \bibinfo {author}
  {\bibfnamefont {F.}~\bibnamefont {Kr\"uger}}, \bibinfo {author}
  {\bibfnamefont {A.~G.}\ \bibnamefont {Green}}, \bibinfo {author}
  {\bibfnamefont {F.}~\bibnamefont {L\'evy-Bertrand}}, \bibinfo {author}
  {\bibfnamefont {B.}~\bibnamefont {Grenier}}, \ and\ \bibinfo {author}
  {\bibfnamefont {A.~D.}\ \bibnamefont {Huxley}},\ }\bibfield  {title}
  {\enquote {\bibinfo {title} {\textit{Modulated magnetism in PrPtAl}},}\
  }\href {\doibase 10.1038/nphys3238} {\bibfield  {journal} {\bibinfo
  {journal} {Nat. Phys.}\ }\textbf {\bibinfo {volume} {11}},\ \bibinfo {pages}
  {321--327} (\bibinfo {year} {2015})}\BibitemShut {NoStop}%
\bibitem [{\citenamefont {Hidaka}\ \emph {et~al.}(2003)\citenamefont {Hidaka},
  \citenamefont {Tokiwa}, \citenamefont {Fujii}, \citenamefont {Watanabe},\
  and\ \citenamefont {Akimitsu}}]{HidakaTokiwa03}%
  \BibitemOpen
  \bibfield  {author} {\bibinfo {author} {\bibfnamefont {M.}~\bibnamefont
  {Hidaka}}, \bibinfo {author} {\bibfnamefont {N.}~\bibnamefont {Tokiwa}},
  \bibinfo {author} {\bibfnamefont {M.}~\bibnamefont {Fujii}}, \bibinfo
  {author} {\bibfnamefont {S.}~\bibnamefont {Watanabe}}, \ and\ \bibinfo
  {author} {\bibfnamefont {J.}~\bibnamefont {Akimitsu}},\ }\bibfield  {title}
  {\enquote {\bibinfo {title} {\textit{Correlation between the structural and
  antiferromagnetic phase transitions in ZnCr$_\text{\!2}$Se$_\text{4}$}},}\
  }\href {\doibase 10.1002/pssb.200301502} {\bibfield  {journal} {\bibinfo
  {journal} {Phys. Stat. Sol. (b)}\ }\textbf {\bibinfo {volume} {236}},\
  \bibinfo {pages} {9--18} (\bibinfo {year} {2003})}\BibitemShut {NoStop}%
\bibitem [{\citenamefont {Kleinberger}\ and\ \citenamefont {{de
  Kouchkovsky}}(1966)}]{KleinbergerKouchkovsky66}%
  \BibitemOpen
  \bibfield  {author} {\bibinfo {author} {\bibfnamefont {R.}~\bibnamefont
  {Kleinberger}}\ and\ \bibinfo {author} {\bibfnamefont {R.}~\bibnamefont {{de
  Kouchkovsky}}},\ }\bibfield  {title} {\enquote {\bibinfo {title}
  {\textit{{\'E}tude radiocristallographique {\`a} basse temp{\'e}rature du
  spinelle ZnCr$_\text{\!2}$Se$_\text{4}$}},}\ }\href@noop {} {\bibfield
  {journal} {\bibinfo  {journal} {C. R. Acad. Sci. Paris Ser. B}\ }\textbf
  {\bibinfo {volume} {262}},\ \bibinfo {pages} {628} (\bibinfo {year}
  {1966})}\BibitemShut {NoStop}%
\bibitem [{\citenamefont {Hemberger}\ \emph {et~al.}(2007)\citenamefont
  {Hemberger}, \citenamefont {von Nidda}, \citenamefont {Tsurkan},\ and\
  \citenamefont {Loidl}}]{HembergerNidda07}%
  \BibitemOpen
  \bibfield  {author} {\bibinfo {author} {\bibfnamefont {J.}~\bibnamefont
  {Hemberger}}, \bibinfo {author} {\bibfnamefont {H.-A.~Krug}\ \bibnamefont
  {von Nidda}}, \bibinfo {author} {\bibfnamefont {V.}~\bibnamefont {Tsurkan}},
  \ and\ \bibinfo {author} {\bibfnamefont {A.}~\bibnamefont {Loidl}},\
  }\bibfield  {title} {\enquote {\bibinfo {title} {\textit{Large
  magnetostriction and negative thermal expansion in the frustrated
  antiferromagnet ZnCr$_\text{\!2}$Se$_\text{4}$}},}\ }\href {\doibase
  10.1103/PhysRevLett.98.147203} {\bibfield  {journal} {\bibinfo  {journal}
  {Phys. Rev. Lett.}\ }\textbf {\bibinfo {volume} {98}},\ \bibinfo {pages}
  {147203} (\bibinfo {year} {2007})}\BibitemShut {NoStop}%
\bibitem [{\citenamefont {Portnichenko}\ \emph {et~al.}(2016)\citenamefont
  {Portnichenko}, \citenamefont {Romh\'anyi}, \citenamefont {Onykiienko},
  \citenamefont {Henschel}, \citenamefont {Schmidt}, \citenamefont {Cameron},
  \citenamefont {Surmach}, \citenamefont {Lim}, \citenamefont {Park},
  \citenamefont {Schneidewind}, \citenamefont {Abernathy}, \citenamefont
  {Rosner}, \citenamefont {van~den Brink},\ and\ \citenamefont
  {Inosov}}]{PortnichenkoRomhanyi16}%
  \BibitemOpen
  \bibfield  {author} {\bibinfo {author} {\bibfnamefont {P.~Y.}\ \bibnamefont
  {Portnichenko}}, \bibinfo {author} {\bibfnamefont {J.}~\bibnamefont
  {Romh\'anyi}}, \bibinfo {author} {\bibfnamefont {Y.~A.}\ \bibnamefont
  {Onykiienko}}, \bibinfo {author} {\bibfnamefont {A.}~\bibnamefont
  {Henschel}}, \bibinfo {author} {\bibfnamefont {M.}~\bibnamefont {Schmidt}},
  \bibinfo {author} {\bibfnamefont {A.~S.}\ \bibnamefont {Cameron}}, \bibinfo
  {author} {\bibfnamefont {M.~A.}\ \bibnamefont {Surmach}}, \bibinfo {author}
  {\bibfnamefont {J.~A.}\ \bibnamefont {Lim}}, \bibinfo {author} {\bibfnamefont
  {J.~T.}\ \bibnamefont {Park}}, \bibinfo {author} {\bibfnamefont
  {A.}~\bibnamefont {Schneidewind}}, \bibinfo {author} {\bibfnamefont {D.~L.}\
  \bibnamefont {Abernathy}}, \bibinfo {author} {\bibfnamefont {H.}~\bibnamefont
  {Rosner}}, \bibinfo {author} {\bibfnamefont {Jeroen}\ \bibnamefont {van~den
  Brink}}, \ and\ \bibinfo {author} {\bibfnamefont {D.~S.}\ \bibnamefont
  {Inosov}},\ }\bibfield  {title} {\enquote {\bibinfo {title} {\textit{Magnon
  spectrum of the helimagnetic insulator Cu$_\text{2}$OSeO$_\text{3}$}},}\
  }\href {\doibase 10.1038/ncomms10725} {\bibfield  {journal} {\bibinfo
  {journal} {Nat. Commun.}\ }\textbf {\bibinfo {volume} {7}},\ \bibinfo {pages}
  {10725} (\bibinfo {year} {2016})}\BibitemShut {NoStop}%
\bibitem [{\citenamefont {Felea}\ \emph {et~al.}(2012)\citenamefont {Felea},
  \citenamefont {Yasin}, \citenamefont {G\"unther}, \citenamefont
  {Deisenhofer}, \citenamefont {Krug~von Nidda}, \citenamefont {Zherlitsyn},
  \citenamefont {Tsurkan}, \citenamefont {Lemmens}, \citenamefont {Wosnitza},\
  and\ \citenamefont {Loidl}}]{FeleaYasin12}%
  \BibitemOpen
  \bibfield  {author} {\bibinfo {author} {\bibfnamefont {V.}~\bibnamefont
  {Felea}}, \bibinfo {author} {\bibfnamefont {S.}~\bibnamefont {Yasin}},
  \bibinfo {author} {\bibfnamefont {A.}~\bibnamefont {G\"unther}}, \bibinfo
  {author} {\bibfnamefont {J.}~\bibnamefont {Deisenhofer}}, \bibinfo {author}
  {\bibfnamefont {H.-A.}\ \bibnamefont {Krug~von Nidda}}, \bibinfo {author}
  {\bibfnamefont {S.}~\bibnamefont {Zherlitsyn}}, \bibinfo {author}
  {\bibfnamefont {V.}~\bibnamefont {Tsurkan}}, \bibinfo {author} {\bibfnamefont
  {P.}~\bibnamefont {Lemmens}}, \bibinfo {author} {\bibfnamefont
  {J.}~\bibnamefont {Wosnitza}}, \ and\ \bibinfo {author} {\bibfnamefont
  {A.}~\bibnamefont {Loidl}},\ }\bibfield  {title} {\enquote {\bibinfo {title}
  {\textit{Spin-lattice coupling in the frustrated antiferromagnet
  ZnCr$_\text{\!2}$Se$_\text{4}$ probed by ultrasound}},}\ }\href {\doibase
  10.1103/PhysRevB.86.104420} {\bibfield  {journal} {\bibinfo  {journal} {Phys.
  Rev.~B}\ }\textbf {\bibinfo {volume} {86}},\ \bibinfo {pages} {104420}
  (\bibinfo {year} {2012})}\BibitemShut {NoStop}%
\bibitem [{\citenamefont {Abernathy}\ \emph {et~al.}(2012)\citenamefont
  {Abernathy}, \citenamefont {Stone}, \citenamefont {Loguillo}, \citenamefont
  {Lucas}, \citenamefont {Delaire}, \citenamefont {Tang}, \citenamefont {Lin},\
  and\ \citenamefont {Fultz}}]{AbernathyStone12}%
  \BibitemOpen
  \bibfield  {author} {\bibinfo {author} {\bibfnamefont {D.~L.}\ \bibnamefont
  {Abernathy}}, \bibinfo {author} {\bibfnamefont {M.~B.}\ \bibnamefont
  {Stone}}, \bibinfo {author} {\bibfnamefont {M.~J.}\ \bibnamefont {Loguillo}},
  \bibinfo {author} {\bibfnamefont {M.~S.}\ \bibnamefont {Lucas}}, \bibinfo
  {author} {\bibfnamefont {O.}~\bibnamefont {Delaire}}, \bibinfo {author}
  {\bibfnamefont {X.}~\bibnamefont {Tang}}, \bibinfo {author} {\bibfnamefont
  {J.~Y.~Y.}\ \bibnamefont {Lin}}, \ and\ \bibinfo {author} {\bibfnamefont
  {B.}~\bibnamefont {Fultz}},\ }\bibfield  {title} {\enquote {\bibinfo {title}
  {\textit{Design and operation of the wide angular-range chopper spectrometer
  ARCS at the Spallation Neutron Source}},}\ }\href {\doibase
  10.1063/1.3680104} {\bibfield  {journal} {\bibinfo  {journal} {Rev. Sci.
  Instrum.}\ }\textbf {\bibinfo {volume} {83}},\ \bibinfo {pages} {015114}
  (\bibinfo {year} {2012})}\BibitemShut {NoStop}%
\bibitem [{\citenamefont {Ollivier}\ and\ \citenamefont
  {Mutka}(2011)}]{OllivierMutka11}%
  \BibitemOpen
  \bibfield  {author} {\bibinfo {author} {\bibfnamefont {J.}~\bibnamefont
  {Ollivier}}\ and\ \bibinfo {author} {\bibfnamefont {H.}~\bibnamefont
  {Mutka}},\ }\bibfield  {title} {\enquote {\bibinfo {title} {\textit{IN5 cold
  neutron time-of-flight spectrometer, prepared to tackle single crystal
  spectroscopy}},}\ }\href {\doibase 10.1143/JPSJS.80SB.SB003} {\bibfield
  {journal} {\bibinfo  {journal} {J.~Phys. Soc. Jpn.}\ }\textbf {\bibinfo
  {volume} {80}},\ \bibinfo {pages} {SB003} (\bibinfo {year}
  {2011})}\BibitemShut {NoStop}%
\bibitem [{\citenamefont {Perring}\ \emph {et~al.}()\citenamefont {Perring},
  \citenamefont {Ewings},\ and\ \citenamefont {Duijn}}]{Horace}%
  \BibitemOpen
  \bibfield  {author} {\bibinfo {author} {\bibfnamefont {T.}~\bibnamefont
  {Perring}}, \bibinfo {author} {\bibfnamefont {R.~A.}\ \bibnamefont {Ewings}},
  \ and\ \bibinfo {author} {\bibfnamefont {J.~V.}\ \bibnamefont {Duijn}},\
  }\href {\doibase http://horace.isis.rl.ac.uk} {}\bibinfo {howpublished}
  {\textsc{Horace} software at
  \href{http://horace.isis.rl.ac.uk}{http://horace.isis.rl.ac.uk}}\BibitemShut
  {NoStop}%
\bibitem [{\citenamefont {Ewings}\ \emph {et~al.}(2016)\citenamefont {Ewings},
  \citenamefont {Buts}, \citenamefont {Le}, \citenamefont {van Duijn},
  \citenamefont {Bustinduy},\ and\ \citenamefont {Perring}}]{EwingsButs16}%
  \BibitemOpen
  \bibfield  {author} {\bibinfo {author} {\bibfnamefont {R.~A.}\ \bibnamefont
  {Ewings}}, \bibinfo {author} {\bibfnamefont {A.}~\bibnamefont {Buts}},
  \bibinfo {author} {\bibfnamefont {M.~D.}\ \bibnamefont {Le}}, \bibinfo
  {author} {\bibfnamefont {J.}~\bibnamefont {van Duijn}}, \bibinfo {author}
  {\bibfnamefont {I.}~\bibnamefont {Bustinduy}}, \ and\ \bibinfo {author}
  {\bibfnamefont {T.~G.}\ \bibnamefont {Perring}},\ }\bibfield  {title}
  {\enquote {\bibinfo {title} {\textit{H{\protect\scalebox{0.8}{ORACE}}:
  Software for the analysis of data from single crystal spectroscopy
  experiments at time-of-flight neutron instruments}},}\ }\href {\doibase
  10.1016/j.nima.2016.07.036} {\bibfield  {journal} {\bibinfo  {journal} {Nucl.
  Instrum. Meth.~A}\ }\textbf {\bibinfo {volume} {834}},\ \bibinfo {pages}
  {132--142} (\bibinfo {year} {2016})}\BibitemShut {NoStop}%
\bibitem [{\citenamefont {Schneidewind}\ and\ \citenamefont
  {\v{C}erm\'{a}k}(2015)}]{SchneidewindCermak15}%
  \BibitemOpen
  \bibfield  {author} {\bibinfo {author} {\bibfnamefont {A.}~\bibnamefont
  {Schneidewind}}\ and\ \bibinfo {author} {\bibfnamefont {P.}~\bibnamefont
  {\v{C}erm\'{a}k}},\ }\bibfield  {title} {\enquote {\bibinfo {title}
  {\textit{P{\protect\scalebox{0.8}{ANDA}}: Cold three axes spectrometer}},}\
  }\href {\doibase 10.17815/jlsrf-1-35} {\bibfield  {journal} {\bibinfo
  {journal} {J. Large-Scale Res. Facilities}\ }\textbf {\bibinfo {volume}
  {1}},\ \bibinfo {pages} {A12} (\bibinfo {year} {2015})}\BibitemShut {NoStop}%
\bibitem [{\citenamefont {Kim}\ \emph {et~al.}(2014)\citenamefont {Kim},
  \citenamefont {Jain}, \citenamefont {Reehuis}, \citenamefont {Khaliullin},
  \citenamefont {Peets}, \citenamefont {Ulrich}, \citenamefont {Park},
  \citenamefont {Faulhaber}, \citenamefont {Hoser}, \citenamefont {Walker},
  \citenamefont {Adroja}, \citenamefont {Walters}, \citenamefont {Inosov},
  \citenamefont {Maljuk},\ and\ \citenamefont {Keimer}}]{KimJain14}%
  \BibitemOpen
  \bibfield  {author} {\bibinfo {author} {\bibfnamefont {J.-H.}\ \bibnamefont
  {Kim}}, \bibinfo {author} {\bibfnamefont {A.}~\bibnamefont {Jain}}, \bibinfo
  {author} {\bibfnamefont {M.}~\bibnamefont {Reehuis}}, \bibinfo {author}
  {\bibfnamefont {G.}~\bibnamefont {Khaliullin}}, \bibinfo {author}
  {\bibfnamefont {D.~C.}\ \bibnamefont {Peets}}, \bibinfo {author}
  {\bibfnamefont {C.}~\bibnamefont {Ulrich}}, \bibinfo {author} {\bibfnamefont
  {J.~T.}\ \bibnamefont {Park}}, \bibinfo {author} {\bibfnamefont
  {E.}~\bibnamefont {Faulhaber}}, \bibinfo {author} {\bibfnamefont
  {A.}~\bibnamefont {Hoser}}, \bibinfo {author} {\bibfnamefont {H.~C.}\
  \bibnamefont {Walker}}, \bibinfo {author} {\bibfnamefont {D.~T.}\
  \bibnamefont {Adroja}}, \bibinfo {author} {\bibfnamefont {A.~C.}\
  \bibnamefont {Walters}}, \bibinfo {author} {\bibfnamefont {D.~S.}\
  \bibnamefont {Inosov}}, \bibinfo {author} {\bibfnamefont {A.}~\bibnamefont
  {Maljuk}}, \ and\ \bibinfo {author} {\bibfnamefont {B.}~\bibnamefont
  {Keimer}},\ }\bibfield  {title} {\enquote {\bibinfo {title}
  {\textit{Competing exchange interactions on the verge of a metal-insulator
  transition in the two-dimensional spiral magnet
  Sr$_\text{\!3}$Fe$_\text{2}$O$_\text{7}$}},}\ }\href {\doibase
  10.1103/PhysRevLett.113.147206} {\bibfield  {journal} {\bibinfo  {journal}
  {Phys. Rev. Lett.}\ }\textbf {\bibinfo {volume} {113}},\ \bibinfo {pages}
  {147206} (\bibinfo {year} {2014})}\BibitemShut {NoStop}%
\bibitem [{\citenamefont {Toth}()}]{SpinW}%
  \BibitemOpen
  \bibfield  {author} {\bibinfo {author} {\bibfnamefont {S.}~\bibnamefont
  {Toth}},\ }\href@noop {} {}\bibinfo {howpublished} {{\textsc{SpinW} software
  at \href{https://www.psi.ch/spinw}{https://www.psi.ch/spinw}}}\BibitemShut
  {NoStop}%
\bibitem [{\citenamefont {Toth}\ and\ \citenamefont {Lake}(2015)}]{TothLake15}%
  \BibitemOpen
  \bibfield  {author} {\bibinfo {author} {\bibfnamefont {S.}~\bibnamefont
  {Toth}}\ and\ \bibinfo {author} {\bibfnamefont {B.}~\bibnamefont {Lake}},\
  }\bibfield  {title} {\enquote {\bibinfo {title} {\textit{Linear spin wave
  theory for single-Q incommensurate magnetic structures}},}\ }\href {\doibase
  10.1088/0953-8984/27/16/166002} {\bibfield  {journal} {\bibinfo  {journal}
  {J.~Phys.: Condens. Matter}\ }\textbf {\bibinfo {volume} {27}},\ \bibinfo
  {pages} {166002} (\bibinfo {year} {2015})}\BibitemShut {NoStop}%
\bibitem [{\citenamefont {Kataoka}(1987)}]{Kataoka87}%
  \BibitemOpen
  \bibfield  {author} {\bibinfo {author} {\bibfnamefont {M.}~\bibnamefont
  {Kataoka}},\ }\bibfield  {title} {\enquote {\bibinfo {title} {\textit{Spin
  waves in systems with long period helical spin density waves due to the
  antisymmetric and symmetric exchange interactions}},}\ }\href {\doibase
  10.1143/JPSJ.56.3635} {\bibfield  {journal} {\bibinfo  {journal} {J.~Phys.
  Soc. Jpn.}\ }\textbf {\bibinfo {volume} {56}},\ \bibinfo {pages} {3635--3647}
  (\bibinfo {year} {1987})}\BibitemShut {NoStop}%
\bibitem [{\citenamefont {Toth}\ \emph {et~al.}(2012)\citenamefont {Toth},
  \citenamefont {Lake}, \citenamefont {Hradil}, \citenamefont {Guidi},
  \citenamefont {Rule}, \citenamefont {Stone},\ and\ \citenamefont
  {Islam}}]{TothLake12}%
  \BibitemOpen
  \bibfield  {author} {\bibinfo {author} {\bibfnamefont {S.}~\bibnamefont
  {Toth}}, \bibinfo {author} {\bibfnamefont {B.}~\bibnamefont {Lake}}, \bibinfo
  {author} {\bibfnamefont {K.}~\bibnamefont {Hradil}}, \bibinfo {author}
  {\bibfnamefont {T.}~\bibnamefont {Guidi}}, \bibinfo {author} {\bibfnamefont
  {K.~C.}\ \bibnamefont {Rule}}, \bibinfo {author} {\bibfnamefont {M.~B.}\
  \bibnamefont {Stone}}, \ and\ \bibinfo {author} {\bibfnamefont {A.~T. M.~N.}\
  \bibnamefont {Islam}},\ }\bibfield  {title} {\enquote {\bibinfo {title}
  {\textit{Magnetic soft modes in the distorted triangular antiferromagnet
  $\alpha$-CaCr$_\text{\!2}$O$_\text{4}$}},}\ }\href {\doibase
  10.1103/PhysRevLett.109.127203} {\bibfield  {journal} {\bibinfo  {journal}
  {Phys. Rev. Lett.}\ }\textbf {\bibinfo {volume} {109}},\ \bibinfo {pages}
  {127203} (\bibinfo {year} {2012})}\BibitemShut {NoStop}%
\bibitem [{\citenamefont {Milstein}\ and\ \citenamefont
  {Sushkov}(2011)}]{MilsteinSushkov11}%
  \BibitemOpen
  \bibfield  {author} {\bibinfo {author} {\bibfnamefont {A.~I.}\ \bibnamefont
  {Milstein}}\ and\ \bibinfo {author} {\bibfnamefont {O.~P.}\ \bibnamefont
  {Sushkov}},\ }\bibfield  {title} {\enquote {\bibinfo {title}
  {\textit{Effective field theories and spin-wave excitations in helical
  magnets}},}\ }\href {\doibase 10.1103/PhysRevB.84.195138} {\bibfield
  {journal} {\bibinfo  {journal} {Phys. Rev.~B}\ }\textbf {\bibinfo {volume}
  {84}},\ \bibinfo {pages} {195138} (\bibinfo {year} {2011})}\BibitemShut
  {NoStop}%
\bibitem [{\citenamefont {Chubukov}\ \emph {et~al.}(1994)\citenamefont
  {Chubukov}, \citenamefont {Sachdev},\ and\ \citenamefont
  {Senthil}}]{ChubukovSachdev94}%
  \BibitemOpen
  \bibfield  {author} {\bibinfo {author} {\bibfnamefont {A.~V.}\ \bibnamefont
  {Chubukov}}, \bibinfo {author} {\bibfnamefont {S.}~\bibnamefont {Sachdev}}, \
  and\ \bibinfo {author} {\bibfnamefont {T.}~\bibnamefont {Senthil}},\
  }\bibfield  {title} {\enquote {\bibinfo {title} {\textit{Large-$S$ expansion
  for quantum antiferromagnets on a triangular lattice}},}\ }\href {\doibase
  10.1088/0953-8984/6/42/019} {\bibfield  {journal} {\bibinfo  {journal}
  {J.~Phys.: Condens. Matter}\ }\textbf {\bibinfo {volume} {6}},\ \bibinfo
  {pages} {8891--8902} (\bibinfo {year} {1994})}\BibitemShut {NoStop}%
\bibitem [{\citenamefont {Chernyshev}\ and\ \citenamefont
  {Zhitomirsky}(2006)}]{ChernyshevZhitomirsky06}%
  \BibitemOpen
  \bibfield  {author} {\bibinfo {author} {\bibfnamefont {A.~L.}\ \bibnamefont
  {Chernyshev}}\ and\ \bibinfo {author} {\bibfnamefont {M.~E.}\ \bibnamefont
  {Zhitomirsky}},\ }\bibfield  {title} {\enquote {\bibinfo {title}
  {\textit{Magnon decay in noncollinear quantum antiferromagnets}},}\ }\href
  {\doibase 10.1103/PhysRevLett.97.207202} {\bibfield  {journal} {\bibinfo
  {journal} {Phys. Rev. Lett.}\ }\textbf {\bibinfo {volume} {97}},\ \bibinfo
  {pages} {207202} (\bibinfo {year} {2006})}\BibitemShut {NoStop}%
\bibitem [{\citenamefont {Chernyshev}\ and\ \citenamefont
  {Zhitomirsky}(2009)}]{ChernyshevZhitomirsky09}%
  \BibitemOpen
  \bibfield  {author} {\bibinfo {author} {\bibfnamefont {A.~L.}\ \bibnamefont
  {Chernyshev}}\ and\ \bibinfo {author} {\bibfnamefont {M.~E.}\ \bibnamefont
  {Zhitomirsky}},\ }\bibfield  {title} {\enquote {\bibinfo {title}
  {\textit{Spin waves in a triangular lattice antiferromagnet: Decays, spectrum
  renormalization, and singularities}},}\ }\href {\doibase
  10.1103/PhysRevB.79.144416} {\bibfield  {journal} {\bibinfo  {journal} {Phys.
  Rev.~B}\ }\textbf {\bibinfo {volume} {79}},\ \bibinfo {pages} {144416}
  (\bibinfo {year} {2009})}\BibitemShut {NoStop}%
\bibitem [{\citenamefont {Dai}(2015)}]{Dai15}%
  \BibitemOpen
  \bibfield  {author} {\bibinfo {author} {\bibfnamefont {P.}~\bibnamefont
  {Dai}},\ }\bibfield  {title} {\enquote {\bibinfo {title}
  {\textit{Antiferromagnetic order and spin dynamics in iron-based
  superconductors}},}\ }\href {\doibase 10.1103/RevModPhys.87.855} {\bibfield
  {journal} {\bibinfo  {journal} {Rev. Mod. Phys.}\ }\textbf {\bibinfo {volume}
  {87}},\ \bibinfo {pages} {855--896} (\bibinfo {year} {2015})}\BibitemShut
  {NoStop}%
\bibitem [{\citenamefont {Inosov}(2016)}]{Inosov16}%
  \BibitemOpen
  \bibfield  {author} {\bibinfo {author} {\bibfnamefont {D.~S.}\ \bibnamefont
  {Inosov}},\ }\bibfield  {title} {\enquote {\bibinfo {title} {\textit{Spin
  fluctuations in iron pnictides and chalcogenides: From antiferromagnetism to
  superconductivity}},}\ }\href {\doibase 10.1016/j.crhy.2015.03.001}
  {\bibfield  {journal} {\bibinfo  {journal} {C. R. Physique}\ }\textbf
  {\bibinfo {volume} {17}},\ \bibinfo {pages} {60--89} (\bibinfo {year}
  {2016})}\BibitemShut {NoStop}%
\bibitem [{\citenamefont {Singh}\ \emph {et~al.}(2003)\citenamefont {Singh},
  \citenamefont {Zheng}, \citenamefont {Oitmaa}, \citenamefont {Sushkov},\ and\
  \citenamefont {Hamer}}]{SinghZheng03}%
  \BibitemOpen
  \bibfield  {author} {\bibinfo {author} {\bibfnamefont {R.~R.~P.}\
  \bibnamefont {Singh}}, \bibinfo {author} {\bibfnamefont {W.}~\bibnamefont
  {Zheng}}, \bibinfo {author} {\bibfnamefont {J.}~\bibnamefont {Oitmaa}},
  \bibinfo {author} {\bibfnamefont {O.~P.}\ \bibnamefont {Sushkov}}, \ and\
  \bibinfo {author} {\bibfnamefont {C.~J.}\ \bibnamefont {Hamer}},\ }\bibfield
  {title} {\enquote {\bibinfo {title} {\textit{Symmetry breaking in the
  collinear phase of the $J_\text{1}$--$J_\text{2}$ Heisenberg model}},}\
  }\href {\doibase 10.1103/PhysRevLett.91.017201} {\bibfield  {journal}
  {\bibinfo  {journal} {Phys. Rev. Lett.}\ }\textbf {\bibinfo {volume} {91}},\
  \bibinfo {pages} {017201} (\bibinfo {year} {2003})}\BibitemShut {NoStop}%
\bibitem [{\citenamefont {Uhrig}\ \emph {et~al.}(2009)\citenamefont {Uhrig},
  \citenamefont {Holt}, \citenamefont {Oitmaa}, \citenamefont {Sushkov},\ and\
  \citenamefont {Singh}}]{UhrigHolt09}%
  \BibitemOpen
  \bibfield  {author} {\bibinfo {author} {\bibfnamefont {G.~S.}\ \bibnamefont
  {Uhrig}}, \bibinfo {author} {\bibfnamefont {M.}~\bibnamefont {Holt}},
  \bibinfo {author} {\bibfnamefont {J.}~\bibnamefont {Oitmaa}}, \bibinfo
  {author} {\bibfnamefont {O.~P.}\ \bibnamefont {Sushkov}}, \ and\ \bibinfo
  {author} {\bibfnamefont {R.~R.~P.}\ \bibnamefont {Singh}},\ }\bibfield
  {title} {\enquote {\bibinfo {title} {\textit{Pnictides as frustrated quantum
  antiferromagnets close to a quantum phase transition}},}\ }\href {\doibase
  10.1103/PhysRevB.79.092416} {\bibfield  {journal} {\bibinfo  {journal} {Phys.
  Rev.~B}\ }\textbf {\bibinfo {volume} {79}},\ \bibinfo {pages} {092416}
  (\bibinfo {year} {2009})}\BibitemShut {NoStop}%
\bibitem [{\citenamefont {Lu}\ \emph {et~al.}(2014)\citenamefont {Lu},
  \citenamefont {Park}, \citenamefont {Zhang}, \citenamefont {Luo},
  \citenamefont {Nevidomskyy}, \citenamefont {Si},\ and\ \citenamefont
  {Dai}}]{LuPark14}%
  \BibitemOpen
  \bibfield  {author} {\bibinfo {author} {\bibfnamefont {X.}~\bibnamefont
  {Lu}}, \bibinfo {author} {\bibfnamefont {J.~T.}\ \bibnamefont {Park}},
  \bibinfo {author} {\bibfnamefont {R.}~\bibnamefont {Zhang}}, \bibinfo
  {author} {\bibfnamefont {H.}~\bibnamefont {Luo}}, \bibinfo {author}
  {\bibfnamefont {A.~H.}\ \bibnamefont {Nevidomskyy}}, \bibinfo {author}
  {\bibfnamefont {Q.}~\bibnamefont {Si}}, \ and\ \bibinfo {author}
  {\bibfnamefont {P.}~\bibnamefont {Dai}},\ }\bibfield  {title} {\enquote
  {\bibinfo {title} {\textit{Nematic spin correlations in the tetragonal state
  of uniaxial-strained BaFe$_{\text{2}-x}$Ni$_x$As$_\text{2}$}},}\ }\href
  {\doibase 10.1126/science.1251853} {\bibfield  {journal} {\bibinfo  {journal}
  {Science}\ }\textbf {\bibinfo {volume} {345}},\ \bibinfo {pages} {657--660}
  (\bibinfo {year} {2014})}\BibitemShut {NoStop}%
\bibitem [{\citenamefont {Bayrakci}\ \emph {et~al.}(2006)\citenamefont
  {Bayrakci}, \citenamefont {Keller}, \citenamefont {Habicht},\ and\
  \citenamefont {Keimer}}]{BayrakciKeller06}%
  \BibitemOpen
  \bibfield  {author} {\bibinfo {author} {\bibfnamefont {S.~P.}\ \bibnamefont
  {Bayrakci}}, \bibinfo {author} {\bibfnamefont {T.}~\bibnamefont {Keller}},
  \bibinfo {author} {\bibfnamefont {K.}~\bibnamefont {Habicht}}, \ and\
  \bibinfo {author} {\bibfnamefont {B.}~\bibnamefont {Keimer}},\ }\bibfield
  {title} {\enquote {\bibinfo {title} {\textit{Spin-wave lifetimes throughout
  the Brillouin zone}},}\ }\href {\doibase 10.1126/science.1127756} {\bibfield
  {journal} {\bibinfo  {journal} {Science}\ }\textbf {\bibinfo {volume}
  {312}},\ \bibinfo {pages} {1926--1929} (\bibinfo {year} {2006})}\BibitemShut
  {NoStop}%
\bibitem [{\citenamefont {Prasai}\ \emph {et~al.}(2017)\citenamefont {Prasai},
  \citenamefont {Trump}, \citenamefont {Marcus}, \citenamefont {Akopyan},
  \citenamefont {Huang}, \citenamefont {McQueen},\ and\ \citenamefont
  {Cohn}}]{PrasaiTrump17}%
  \BibitemOpen
  \bibfield  {author} {\bibinfo {author} {\bibfnamefont {N.}~\bibnamefont
  {Prasai}}, \bibinfo {author} {\bibfnamefont {B.~A.}\ \bibnamefont {Trump}},
  \bibinfo {author} {\bibfnamefont {G.~G.}\ \bibnamefont {Marcus}}, \bibinfo
  {author} {\bibfnamefont {A.}~\bibnamefont {Akopyan}}, \bibinfo {author}
  {\bibfnamefont {S.~X.}\ \bibnamefont {Huang}}, \bibinfo {author}
  {\bibfnamefont {T.~M.}\ \bibnamefont {McQueen}}, \ and\ \bibinfo {author}
  {\bibfnamefont {J.~L.}\ \bibnamefont {Cohn}},\ }\bibfield  {title} {\enquote
  {\bibinfo {title} {\textit{Ballistic magnon heat conduction and possible
  Poiseuille flow in the helimagnetic insulator
  Cu$_\text{2}$OSeO$_\text{3}$}},}\ }\href {\doibase
  10.1103/PhysRevB.95.224407} {\bibfield  {journal} {\bibinfo  {journal} {Phys.
  Rev.~B}\ }\textbf {\bibinfo {volume} {95}},\ \bibinfo {pages} {224407}
  (\bibinfo {year} {2017})}\BibitemShut {NoStop}%
\end{thebibliography}%

\end{document}